\documentclass[usenatbib,usenatbib]{mnras}
\usepackage{newtxtext,newtxmath}
\usepackage{todonotes}
\usepackage{times}
\usepackage{amssymb}
\usepackage{amsmath}
\usepackage{graphicx} 
\usepackage{bmpsize}
\usepackage{epstopdf}
\usepackage{dblfloatfix} 
\graphicspath{{./image/}}
\usepackage[normalem]{ulem}
\usepackage[T1]{fontenc}
\usepackage{makecell}

\newcommand{\HI}{\rm H~{\sc i }}
\newcommand{\HII}{\rm H~{\sc ii }}

\newcommand{\TB}{\delta T_{\rm b}}
\newcommand{\MSUN}{{\rm M}_{\odot}}

\newcommand{\XHI}{x_{\rm HI}}
\newcommand{\XHII}{x_{\rm HII}}
\newcommand{\TS}{T_{\rm S}}
\newcommand{\TK}{T_{\rm K}}
\newcommand{\TCMB}{T_{\gamma}}
\newcommand{\lya}{\rm {Ly{\alpha}}}
\newcommand{\OmegaB}{\Omega_{\rm B}}
\newcommand{\Omegam}{\Omega_{\rm m}}

\title[Constraining the intergalactic medium at $z\approx$ 9.1 with the LOFAR]{Constraining the intergalactic medium at $z\approx$ 9.1 using LOFAR Epoch of Reionization observations}

\author[Ghara et al.]{R. Ghara$^{1, 2, 3}$\thanks{E-mail: ghara.raghunath@gmail.com}, S. K. Giri$^{1,4}$\thanks{E-mail: sambit.giri@gmail.com}, G. Mellema$^{1}$, B. Ciardi$^{5}$, S. Zaroubi$^{2,3,6}$, I. T. Iliev$^{7}$,  
\newauthor  L. V. E. Koopmans$^{6}$, E. Chapman$^{8}$, S. Gazagnes$^{6}$, B. K. Gehlot$^{9,6}$, A. Ghosh$^{10,11,18}$, V. Jeli\'c$^{12}$,  
\newauthor
F. G. Mertens$^{6,13}$, R. Mondal$^{7}$, J. Schaye$^{14}$, M. B. Silva$^{15}$, K. M. B. Asad$^{16}$,  R. Kooistra$^{6, 19}$, 
\newauthor
M. Mevius$^{17}$, A. R. Offringa$^{6, 17}$, V. N. Pandey$^{6,17}$ and S. Yatawatta$^{17}$
\\
$^{1}$The Oskar Klein Centre, Department of Astronomy, Stockholm University, AlbaNova, SE-10691 Stockholm, Sweden\\
$^{2}$Department of Natural Sciences, The Open University of Israel, 1 University Road, PO Box 808, Ra'anana 4353701, Israel \\
$^{3}$Department of Physics, Technion, Haifa 32000, Israel\\
$^{4}$Institute for Computational Science, University of Zurich, Winterthurerstrasse 190, 8057 Zurich, Switzerland \\
$^{5}$Max-Planck Institute for Astrophysics, Karl-Schwarzschild-Stra{\ss}e 1, 85748 Garching, Germany\\
$^{6}$Kapteyn Astronomical Institute, University of Groningen, PO Box 800, 9700AV Groningen, the Netherlands\\
$^{7}$Astronomy Centre, Department of Physics and Astronomy, Pevensey II Building, University of Sussex, Brighton BN1 9QH, U.K.\\
$^{8}$Astrophysics Group, Imperial College London, Blackett Laboratory, Prince Consort Road, London, SW7 2AZ, United Kingdom\\
$^{9}$School of Earth and Space Exploration, Arizona State University, Tempe, AZ, United States \\
$^{10}$Department of Physics, University of the Western Cape, Cape Town 7535, South Africa\\
$^{11}$SARAO, 2 Fir Street, Black River Park, Observatory, Capetown, South Africa\\
$^{12}$Ru{\dj}er Bo\v{s}kovi\'{c} Institute, Bijeni\v{c}ka cesta 54, 10000 Zagreb, Croatia\\
$^{13}$LERMA, Observatoire de Paris, PSL Research University, CNRS, Sorbonne Université, F-75014 Paris, France\\
$^{14}$Leiden Observatory, Leiden University, PO Box 9513, 2300RA Leiden, the Netherlands \\
$^{15}$Institute of Theoretical Astrophysics, University of Oslo, PO Box 1029 Blindern, N-0315 Oslo, Norway\\
$^{16}$Independent University Bangladesh, Plot 16, Block B, Aftabuddin Ahmed Road, Bashundhara R/A, Dhaka, Bangladesh \\
$^{17}$ Astron, PO Box 2, 7990 AA Dwingeloo, the Netherlands \\
$^{18}$ Department of Physics, Banwarilal Bhalotia College, Asansol, West Bengal, India\\
$^{19}$ Kavli IPMU (WPI), UTIAS, The University of Tokyo, Kashiwa, Chiba 277-8583, Japan\\
}

\date{Accepted XXX. Received YYY; in original form ZZZ}

\pubyear{2020}

\begin{document}
\label{firstpage}
\pagerange{\pageref{firstpage}--\pageref{lastpage}}
\maketitle
\begin{abstract}
We derive constraints on the thermal and ionization states of the intergalactic medium (IGM) at redshift $\approx$ 9.1 using new upper limits on the 21-cm power spectrum measured by the LOFAR radio-telescope and a prior on the ionized fraction at that redshift estimated from recent cosmic microwave background (CMB) observations. We have used results from the reionization simulation code {\sc grizzly} and a Bayesian inference framework to constrain the parameters which describe the physical state of the IGM. We find that, if the gas heating remains negligible, an IGM with ionized fraction $\gtrsim 0.13$ and a distribution of the ionized regions with a characteristic size $\gtrsim 8 ~h^{-1}$ comoving megaparsec (Mpc) and a full width at the half maximum (FWHM) $\gtrsim 16 ~h^{-1}$ Mpc is ruled out.  For an IGM with a uniform spin temperature $\TS \gtrsim 3$ K, no constraints on the ionized component can be computed. If the large-scale fluctuations of the signal are driven by spin temperature fluctuations, an IGM with a volume fraction $\lesssim 0.34$ of heated regions with a temperature larger than CMB, average gas temperature 7-160 K and a distribution of the heated regions with characteristic size 3.5-70 $h^{-1}$ Mpc and FWHM of $\lesssim 110$ $h^{-1}$ Mpc is ruled out. These constraints are within the 95 per cent credible intervals. With more stringent future upper limits from LOFAR at multiple redshifts, the constraints will become tighter and will exclude an increasingly large region of the parameter space.

\end{abstract}


\section{ Introduction} 
\label{sec:intro}

The Epoch of Reionization (EoR) is one of the least understood chapters in the history of our Universe. The formation of the first luminous sources initiated the transition of the cold and neutral intergalactic medium (IGM) into a hot and ionized state. This transition had a significant impact on the later stages of structure formation through various feedback mechanisms (see e.g. \citealt{Ciardi2005} for a review). Although we know that reionization took place, very few facts about it are known with certainty \citep[see e.g.][for reviews]{Morales10, Pritchard12, 2013ASSL..396...45Z, 2016PhR...645....1B}. 

Theoretical models suggest that ionizing ultra-violet (UV) photons from the first sources created localized ionized regions, which over time grew in size, started to overlap and, as an increasing number of sources formed, led to a complete reionization of the IGM. Observations of high-redshift ($z\gtrsim6$) quasar absorption spectra suggest that complete reionization was reached around redshift 6 \citep[e.g.][]{Fan06b, Mortlock11, Venemans15,  2018Natur.553..473B}. On the other hand, the measurement of the Thomson optical depth from the observation of Cosmic Microwave Background (CMB) \citep{2018arXiv180706209P} suggests that the probable period of this event lies at redshift $\lesssim$10  \citep{ Choudhury06a, mitra2011, mitra2012}.  However, the details of the reionization process such as the exact timing of the EoR, the morphology of the \HI distribution in the IGM and the properties of early sources, are still poorly known.

The redshifted 21-cm signal from neutral hydrogen in the IGM is the most promising probe of the EoR, as it has the ability to reveal many of the unknown facts about this epoch. Inspired by its potential, several radio telescopes such as the Low Frequency Array (LOFAR)\footnote{\url{http://www.lofar.org/}} \citep{vanHaarlem2013LOFAR:ARray,2017ApJ...838...65P}, 
the Precision Array for Probing the Epoch of Reionization (PAPER)\footnote{\url{http://eor.berkeley.edu/}} \citep{parsons13, 2019ApJ...883..133K}, the Murchison Widefield Array (MWA)\footnote{\url{http://www.mwatelescope.org/}} \citep{bowman13, 2019ApJ...884....1B} and the Hydrogen Epoch of Reionization Array (HERA)\footnote{\url{https://reionization.org/}} \citep{2017PASP..129d5001D} have invested considerable amounts of observing time to detect this signal. Due to their limited sensitivity, these radio interferometers aim to measure the statistical fluctuations of the signal. The planned Square Kilometre Array (SKA)\footnote{\url{http://www.skatelescope.org/}} will in addition be able to produce actual tomographic images of the distribution of the signal on the sky \citep{2015aska.confE..10M, ghara16}. Beside these large radio interferometers, single antenna experiments such as EDGES \citep{2010Natur.468..796B}, EDGES2 \citep{monsalve2017,EDGES2018}, SARAS \citep{2015ApJ...801..138P}, SARAS2 \citep{singh2017}, BigHorns \citep{2015PASA...32....4S}, SciHi  \citep{2014ApJ...782L...9V} and LEDA \citep{price2018}  are being used to attempt a detection of the sky-averaged 21-cm signal and its evolution with redshift.

In spite of all these efforts, so far no undisputed detection of the 21-cm signal from the EoR has been made. The main reason for this is that the signal is several orders of magnitude weaker than the galactic and extra-galactic foregrounds at these frequencies \citep[see e.g.,][]{Shaver1999A&A.345.380S, jelic08}. Moreover, the signal low amplitude implies long integration times are required to exceed the instrumental noise. Although there exist accurate methods to subtract \citep{2009MNRAS.397.1138H, 2015MNRAS.447.1973B, 2013MNRAS.429..165C, 2016MNRAS.458.2928C, 2018MNRAS.478.3640M}, suppress \citep{kanan2007MNRAS.382..809D, 2012MNRAS.426.3178M, ghara15c} or avoid \citep{2010ApJ...724..526D, 2014PhRvD..90b3019L} the foregrounds, these only work if the sky signal has been measured with high fidelity over the time of observation. This then requires exquisite calibration of the system as any left-over artefacts from strong sources will make a measurement impossible \citep{2016MNRAS.461.3135B, 2017ApJ...838...65P}. This implies calibrating the many hardware components of the telescope \citep[see e.g.,][]{2019ApJ...884..105K} while a further complication is added by the presence of the temporally and spatially varying ionosphere \citep[see e.g.,][]{2016RaSc...51..927M}.

Recently, \citet{EDGES2018} have claimed a detection of the sky-averaged 21-cm signal at $z\approx 17$ in observations with the EDGES2 low-band antenna. These results are debated \citep[e.g.\ in][] {2018Natur.564E..32H, 2018ApJ...858L..10D, 2019ApJ...880...26S, 2019ApJ...874..153B}, but if true would challenge our theoretical understanding of the early universe as explanations for its strength require either a previously unknown cooling mechanism \citep[see e.g.,][]{2014PhRvD..90h3522T, 2018Natur.555...71B, 2018PhRvL.121a1101F, 2018arXiv180210094M} or a radio background other than the CMB \citep[][]{2018ApJ...858L..17F, 2019MNRAS.486.1763F}.

Other attempts have to date only provided upper limits on the expected signal. While global signal experiments probe the average brightness temperature, experiments with radio interferometers constrain the power spectrum of the expected 21-cm signal. Observations with the GMRT \citep{paciga13} provided the very first upper limit, which was a 2-sigma value of $(248)^2~ \rm mK^2$ for $k=0.50 ~h ~\rm Mpc^{-1}$ at $z=8.6$. Later PAPER and MWA produced additional upper limits \citep{parsons13, 2015ApJ...809...61A,  2019ApJ...884....1B}. Note that the PAPER collaboration initially reported a strong upper bound \citep{2016ApJ...833..102B} which was later revised to a weaker upper bound \citep{2018ApJ...868...26C, 2019ApJ...883..133K}.  The first LOFAR upper limit on the power spectrum of the signal obtained from one night observation was $(79.6)^2 \rm ~mK^2$ at $k = 0.053 ~h ~\rm Mpc^{-1}$ and a redshift between 9.6 and 10.6  \citep{2017ApJ...838...65P}.
Recently, upper limits were provided for even higher redshifts.  \citet{2019MNRAS.488.4271G} placed upper limits on the power spectrum in the redshift range $z = 19.8-25.2$ using observations with the LOFAR-Low Band Antenna array and \citet{2019AJ....158...84E} placed upper limits at $z\approx 18.4$ using observations with the Owens Valley Radio Observatory Long Wavelength Array (OVRO-LWA)\footnote{\url{http://www.tauceti.caltech.edu/LWA/}}.

\citet{2019LOFAR} have provided the second LOFAR upper limit on the 21-cm power spectrum at redshift $\approx 9.1$ using 10 nights of observations. At $k = 0.1 ~h ~\rm Mpc^{-1}$ the $2-\sigma$ upper limit is $(106.65)^2 \rm ~mK^2$, a factor of $\approx 8$ improvement at the same $k-$ scale over the value obtained from 1 night of observations \citep{2017ApJ...838...65P} and the best upper limit so far on the large-scale power spectrum at redshift 9. The results give upper limits for a range of $k$ values but only at one redshift, $z\approx 9.1$. In this paper we explore scenarios for the EoR that can be ruled out by these new upper limits. As they are about an order of magnitude higher than the most popular theoretical predictions, we can expect that only extreme models will be ruled out. Similar analyses was performed by \citet{2015ApJ...809...62P} and \citet{2016MNRAS.455.4295G} for the earlier upper limits from PAPER which were reported in \citet{2015ApJ...809...61A}.

Extracting astrophysical and cosmological information from 21-cm observations is not straightforward as, in addition to the cosmological dependence, the characteristics of the expected signal depend crucially on specific properties of the early sources and their redshift evolution. While UV photons from such sources are mostly absorbed during ionization of \HI in surrounding regions, X-ray photons, due to their longer mean free path, penetrate further into the neutral gas and increase its temperature \citep[see e.g.,][]{madau1997, 2005MNRAS.360L..64Z, 2007MNRAS.375.1269Z}. At the same time, $\lya$ photons from the same sources  determine the coupling strength of the \HI spin temperature with the gas temperature. In view of this complexity, an exploration of many theoretical models of the expected 21-cm signal is necessary to interpret the results from radio observations. Such signal prediction algorithms are often combined with a Bayesian inference framework, such as the Monte Carlo Markov Chains (MCMC), to explore and constrain the reionization parameters \citep[e.g.][]{Greig201521CMMC:Signal,2017MNRAS.472.2651G,Park2019InferringSignal, 2019arXiv191006274C}. This is the approach we will use here, relying on the {\sc grizzly} code \citep{ghara15a, ghara18} to generate the reionization scenarios and models for the 21-cm signal.

Since the codes which are used to generate the 21-cm signal use source parameters as input, the results from such Bayesian inference frameworks typically constrain these source parameters. However, it should be realised that the 21-cm observations themselves characterise the state of the IGM and do not contain any direct information about the source properties. It is perfectly possible that many different source models lead to the same 21-cm signal, especially if one only has information from a single redshift, as is the case for the latest LOFAR upper limits. As explained in more detail below, we will therefore have a strong focus on IGM parameters such as the average ionized fraction, average spin temperature, volume fraction of ``heated region'' (i.e, partially ionized regions with gas temperature larger than the CMB temperature) and size distributions of ionized and heated regions. We give much less weight to the source parameters, which however are still needed by the models to generate the 21-cm signals.

Our paper is structured as follows. In section \ref{sec:method}, we describe the basic methodology to prepare the Bayesian framework used to interpret the observations.  We present our results in section \ref{sec:results} and discuss them from the point of view of other observations in section \ref{sec:discussion}, before concluding in section \ref{sec:con}.
The cosmological parameters as used in this study are  $\Omegam=0.27$, $\Omega_\Lambda=0.73$, $\OmegaB=0.044$, $h=0.7$, consistent with the Wilkinson ~Microwave ~Anisotropy ~Probe (WMAP) results \citep[][]{2013ApJS..208...19H}. These are the same as used in the $N$-body simulation used in this paper. Within the error bars, these  are consistent with Planck 2015 results \citep{2016A&A...594A..13P} which are used in \citet{2019LOFAR}. Note that all distances and scales used in this study are in comoving coordinates.

\section{Methodology}
\label{sec:method}

Here we introduce the approach employed to generate the 21-cm signal and compare it with the observational upper limit.

\subsection{Cosmological simulations}
\label{sec:nbody}
We use the {\sc grizzly} code \citep{ghara15a, ghara18} to generate brightness temperature maps at redshift $\approx 9.1$. This algorithm requires cosmological density fields and halo catalogues as input. These are retrieved from results of the PRACE\footnote{Partnership for Advanced Computing in Europe: \url{http://www.prace-ri.eu/}} project PRACE4LOFAR, which was run specifically for the purpose of providing several cosmological simulations for the interpretation of LOFAR data. Here we use the largest volume, of length 500~$h^{-1}$~ comoving megaparsec (Mpc) \citep[see e.g,][]{Giri2019Position-dependentReionization,Giri2019NeutralTomography}. This corresponds to a field-of-view of $4.27^{\circ}\times 4.27^{\circ}$ at redshift $\approx$ 9.1 which is comparable to the LOFAR primary beam of $\approx 4^{\circ}$. The cosmological $N$-body simulation was run using {\sc cubep$^3$m} \citep{Harnois12} with 6912$^3$ particles and a mass resolution of  $4.05\times 10^7 ~\MSUN$. Halos were identified on the fly with a spherical overdensity halo finder \citep{Watson2013TheAges}, and only those with masses $10^9$~M$_\odot$ and higher, i.e. resolved with at least $\approx 25$ particles, were used.  
More details on the PRACE4LOFAR simulations can be found in \citet{Dixon2016TheReionization}. 

The {\sc grizzly} simulations are run on gridded versions of the density fields  from which the halos have been removed as they are not part of the IGM and their effect is captured by the assumptions of the source model 
through the photon escape fraction. 
We use $300^3$ grids for the results in this paper. The smallest $k$-scale which can be probed with this resolution is $\approx 1.9 ~h ~\rm Mpc^{-1}$ (corresponding to scale $\approx 3.3 ~ h^{-1}~\rm Mpc$). The smallest scale probed in \citet{2019LOFAR}, which is $\approx 0.4 ~h ~\rm Mpc^{-1}$ (corresponding to scale $\approx 15 ~ h^{-1}~\rm Mpc$), remains within the Nyquist limit of our simulation and free from the aliasing effect \citep{Mao2012Redshift-spaceRe-examined}.

\subsection{Modelling the 21-cm signal using {\sc grizzly}}
\label{sec:grizzly}
The differential brightness temperature, $\TB$, of the 21-cm signal can be expressed as \citep[see e.g,][]{madau1997, Furlanetto2006}, 
\begin{eqnarray}
 \TB (\mathbf{x}, z) \!  & = & \! 27 ~ x_{\rm HI} (\mathbf{x}, z) [1+\delta_{\rm B}(\mathbf{x}, z)] \left(\frac{\OmegaB h^2}{0.023}\right) \nonumber\\
&\times& \!\left(\frac{0.15}{\Omegam h^2}\frac{1+z}{10}\right)^{1/2}  \left(1-\frac{\TCMB}{\TS(\mathbf{x}, z)} \right)\,\rm{mK},
\nonumber \\
\label{eq:brightnessT}
\end{eqnarray}
where the quantities  $x_{\rm HI}$, $\delta_{\rm B}$ and $\TCMB(z)$ = $2.73~(1+z)$ K denote the neutral hydrogen fraction, baryonic density contrast and the CMB temperature, respectively, each at position $\mathbf{x}$ and redshift $z$. $\TS$ represents the spin temperature of hydrogen in the IGM. In this paper, we will consider the dimensionless power spectrum of the brightness temperature, i.e., $\Delta^2(k) = k^3 P(k)/2\pi^2$. The spherically averaged power spectrum $P(k)$ can be expressed as $\langle \hat{\TB}(\mathrm{\bf k}) \hat{\TB}^{\star}(\mathbf{k'})\rangle = (2 \pi)^3 \delta_{\rm D}(\mathbf{k - k'}) P(k)$ where $\hat{\TB}(\mathrm{\bf k})$ denotes the Fourier component of $\TB(\mathbf{x})$ at wavenumber $k$.

The {\sc grizzly} algorithm is based on a one-dimensional radiative transfer scheme and is an independent implementation of the {\sc bears} algorithm described by \citet{Thom08,Thom09,Thom11} and \citet{2018NewA...64....9K}. It approximates the transfer of photons by assuming that the effect from individual sources is isotropic and can therefore be pre-calculated as radial profiles around each source. The algorithm corrects for overlap by ensuring that the total ionized volume of the region created by multiple sources is the correct one. This approach makes the code very fast, a requirement necessary  for parameter studies such as the one we perform here.

\citet{ghara18} presented a  detailed comparison between the performance of this code and the full three-dimensional radiative transfer code C$^2${\sc ray} \citep{mellema06}. We found that although {\sc grizzly} employs a range of approximations, its results agree with those of the full radiative transfer quite well, while being at least $10^5$ times faster. In Appendix \ref{app:grizzly}, we give a brief outline of this code, while we refer the reader to the original papers \citep{ghara15a, ghara18} for a more detailed and complete description of the algorithm. Note that we have not included redshift space distortions while evaluating the power spectrum for different model parameters, as their impact remains rather small during the EoR, when ionization fluctuations dominate the power spectrum of $\TB$ \citep{Jensen13, 2016JApA...37...32M, ghara15a}.

\begin{table}
\centering
\caption[]{The $\Delta^2_\mathrm{21}$ upper limits at $1-\sigma$ level at redshift 9.1 from LOFAR observations for different k-bins \citep{2019LOFAR}. }
\small
\tabcolsep 8pt
\renewcommand\arraystretch{1.5}
   \begin{tabular}{c c c c}
\hline
\hline
k ($h ~\rm Mpc^{-1}$) & $\Delta^2_\mathrm{21}(k)$ (mK$^2$) & $\Delta^2_\mathrm{21,err}$ (mK$^2$) \\

\hline
\hline
0.075 & $(58.97)^2$  & $(30.26)^2$ & \\
0.100 & $(95.21)^2$ & $(33.98)^2$  & \\
0.133 & $(142.17)^2$ & $(39.98)^2$ & \\
0.179 & $(235.80)^2$ & $(51.81)^2$  & \\
0.238 & $(358.95)^2$ & $(64.00)^2$  & \\
0.319 & $(505.26)^2$ & $(87.90)^2$  & \\
0.432 & $(664.23)^2$ & $(113.04)^2$  & \\
\hline
\hline
\end{tabular}
\label{tab_obs}
\end{table}

\begin{table*}
\centering
\caption[]{Overview of the source parameters used in {\sc grizzly}, their explored ranges as well as for which models these are used as input parameters.}
\small
\tabcolsep 2pt
\renewcommand\arraystretch{1.5}
   \begin{tabular}{c c c c c}
\hline
\hline
Source Parameters & Description & Explored range & Corresponding Models  	 \\
\hline
\hline
$\zeta$ & Ionization efficiency   & [$10^{-2}, 10^{2.5}$]   &  Varied in both the uniform and non-uniform $\TS$ models     & 	\\
$M_{\rm min}$ & Minimum mass of the UV emitting halos   & [$10^{9}  ~\MSUN, 10^{12} ~\MSUN$]   &  \makecell{Varied in the uniform $\TS$ model\\ Fixed to $10^9 ~\MSUN$ in the non-uniform $\TS$ model}     & 	\\
$M_{\rm min,X}$ & Minimum mass of the X-ray emitting halos   & [$10^{9} ~\MSUN, 10^{12} ~\MSUN$]   &  Used and varied only for the non-uniform $\TS$ model     & 	\\
$f_X$ & X-ray heating efficiency   & [0.1, 10]   &  Used and varied only for the non-uniform $\TS$ model     & 	\\
$\alpha$ & Spectral index of the X-ray spectrum   & Fixed to 1.2 (fiducial) or 0.3   &  Used only for the non-uniform $\TS$ model      & 	\\
\hline
\hline
\end{tabular}
\label{tab_source_param}
\end{table*}

The upper limits from \citet{2019LOFAR} at scales $k=0.075$ and $0.1 ~h$ Mpc$^{-1}$ are $\Delta^2 = (58.97)^2 ~\rm mK^2$ and $(95.21)^2$~mK$^2$ respectively (see also Table \ref{tab_obs}). Before proceeding, it should be realized that these values are rather high compared to the power spectrum at redshift $\approx 9.1$ predicted by various standard reionization scenarios, such as in \citet{2006NewA...11..374M}, \citet{2007MNRAS.376..534I}, \citet{Greig201521CMMC:Signal}, \citet{ghara15b},
\citet{2016MNRAS.457.1550H}, \citet{2018MNRAS.478.5564B} and \citet{2019MNRAS.tmp.1183R}. For example, the predicted power spectra at $z\approx 9$ at $k=0.1 ~h$ Mpc$^{-1}$ are found to be $\lesssim$ $10^3$ mK$^2$. Models which can be excluded by these upper limits therefore have to be quite extreme.

As the lowest upper limit is for the largest scales, our aim is to identify scenarios which produce large amplitudes for the large-scale fluctuations. Spatial fluctuations in the 21-cm signal can only be caused by spatial fluctuations in $x_{\rm HI}$, $\delta_\mathrm{B}$ or $\TS$ (see Eq.~\ref{eq:brightnessT}). Previous studies have shown that the fluctuations in $\delta_\mathrm{B}$ are small on the scales measured by LOFAR \citep[e.g.][]{peebles1993principles}.
We therefore consider two different scenarios to identify models with either large $x_{\rm HI}$ and/or $\TS$ fluctuations. In the first scenario, we assume a uniform $\TS$, so that the large-scale fluctuations of the signal are mostly driven by fluctuations in $x_{\rm HI}$. In the second scenario, we relax the uniform $\TS$ assumption and allow sources of heating to create local regions of high $\TS$. In this case the large-scale fluctuations are predominantly sourced by fluctuations in $\TS$. These two scenarios will be discussed in detail later in Section \ref{sec:coldigm} and ~\ref{sec:heat}.

To calculate the evolution of the IGM for these scenarios, {\sc grizzly} needs to characterize the source properties with a range of parameters. The following  are used in our study (also listed in Table \ref{tab_source_param}).

\begin{itemize}
    \item Ionization efficiency ($\zeta$): the rate of ionizing photons per unit stellar mass escaping from a halo is given by $\dot N_i=\zeta\times 2.85\times 10^{45}  ~{\rm s^{-1}} ~\MSUN^{-1}$. This value corresponding to $\zeta=1$ derives from the model galaxy spectrum employed when calculating $\XHII$ and $\TK$, which has been produced with the publicly available code {\sc pegase2} \footnote{\tt \url{http://www2.iap.fr/pegase/}} \citep{Fioc97}. Note that the emission rate of the ionizing photons is assumed to be proportional to the halo mass. We refer the reader to \citet{ghara15a} for more details. We calculate the stellar mass of a halo using $M_\star = f_\star \times \frac{\OmegaB}{\Omegam} \times M_{\rm halo}$, where $f_\star$ is the star formation efficiency, fixed at 0.02 \citep{2015ApJ...799...32B, 2016MNRAS.460..417S}. The parameter $\zeta$ combines all the degeneracies from various quantities related to the star formation rate and the emission rate of ionizing photons from the sources, as well as their escape fraction into the IGM. The case $\zeta=1$ corresponds to a star formation efficiency of 2 per cent and an escape efficiency of 100 per cent, but also to a star formation efficiency of 20 per cent and an escape efficiency of 10 per cent. We vary $\zeta$ in both scenarios considered in this paper.

    \item Minimum mass of the UV emitting halos ($M_{\rm min}$): In the above parametrization of the ionizing efficiency the number of ionizing photons escaping from a halo depends linearly on its mass. However, below a certain minimum mass radiative and mechanical feedback can severely reduce the star formation efficiency \citep[see e.g.,][]{2013MNRAS.428..154H, 2018MNRAS.480.1740D}. We model this by introducing $M_{\rm min}$ as the minimum mass of halos from which ionizing photons escape into the IGM. As with $\zeta$, this parameter represents different physical processes, not only feedback but for example also very low escape fractions from lower-mass halos 
    \citep[see e.g.,][]{2008ApJ...672..765G, 2016MNRAS.458L..94S}. Due to the mass resolution of our N-body simulation (see Sect.~\ref{sec:nbody}) the lowest value for $M_{\rm min}$ is $10^9$~$\MSUN$. Although halos of lower masses could contribute, as we will see below, the LOFAR results are not able to constrain such low values. In general, one expects star formation in halos with mass $\lesssim 10^9 ~\MSUN$ to be suppressed due to radiative feedback \citep[see e.g.,][]{2014MNRAS.442.2560W, Dixon2016TheReionization}. Note that we do not employ radiative feedback in this study as $M_{\rm min}$ remains $\gtrsim 10^9 ~\MSUN$ for the scenarios considered here. We vary $M_{\rm min}$ in the uniform $\TS$ scenario while fix it to $10^9 ~\MSUN$ in the non-uniform $\TS$ model.

    \item Minimum mass of X-ray emitting halo ($M_{\rm min, X}$): In addition to the stellar contributions, {\sc grizzly} can also include heating and ionization from X-ray sources such as quasars, high-mass X-ray binaries, etc. As not all star hosting halos are necessarily substantial X-ray sources, we use the minimum mass of dark matter halos that contain X-ray sources as a separate parameter. This allows us to include scenarios in which the X-ray source population deviates from the population of galaxies. We consider and vary this parameter only in the non-uniform $\TS$ model.
    
    \item X-ray heating efficiency ($f_X$) and spectral index ($\alpha$):  The spectrum of an X-ray source at energy $E$ is modelled as a power-law, i.e. $I_q(E) = A_q ~ E^{-\alpha}$, where $\alpha$ is the spectral index. The normalization constant $A_q$ is determined such that the X-ray luminosity per stellar mass is $3.4\times 10^{34} f_X ~\rm erg ~s^{-1} ~\MSUN^{-1}$, where $f_X$ is a parameter. This implies a rate of X-ray photons per unit stellar mass emitted from a halo $\dot N_x = f_X \times 8.47 \times 10^{43} ~\rm s^{-1} ~\MSUN^{-1}$. The value of $\dot N_x$ for $f_X=1$ is $\sim$ two orders of magnitude larger than the measurements of high-mass X-ray binaries in local star forming galaxies in 0.5-8 keV band \citep{2012MNRAS.419.2095M}.  We assume that the UV band spans the range 13.6--100 eV, while the X-ray band goes from 100 eV to 10 keV. We vary $f_X$ while we keep $\alpha$ fixed at 1.2 (fiducial) or 0.3 in the non-uniform $\TS$ model.

\end{itemize}

\begin{table*}
\centering
\caption[]{An overview of the IGM parameters considered in this paper. Except for $\TS$ in the case of the uniform $\TS$ model, all of these are {\it derived} from the simulation results. We explore a range [-12:1] for $1-\TCMB/\TS$. The last column refers to the models in which such a quantity is considered.}
\small
\tabcolsep 8pt
\renewcommand\arraystretch{1.5}
   \begin{tabular}{c c c c c}
\hline
\hline
IGM Parameters & Description &  Corresponding Models  	 \\
\hline
\hline
$\overline{\XHII}$ & Volume averaged ionized fraction     &  \makecell{Uniform \\ and non-uniform $\TS$ models}     & 	\\
$\overline{\TK}$ & \makecell{Volume averaged gas temperature in \\ the partially ionized IGM with $\XHII<0.5$}     &  Non-uniform $\TS$ model     & 	\\
$1-\TCMB/\TS$ & \makecell{$\TCMB$ and $\TS$ are the CMB and \\ spin temperature}     &  Uniform $\TS$ model     & 	\\
$\overline{\TB}$ & Mass averaged differential brightness temperature      &  Uniform and non-uniform $\TS$ models    & 	\\
$f_{\rm heat}$ & Volume fraction of regions with temperature larger than $\TCMB$    &  Non-uniform $\TS$ model & \\
$R^{\rm HII}_{\rm peak}$ & \makecell{Size at which the PDF of the size \\ distribution of the \HII regions peaks}   &  Uniform $\TS$ model      & 	\\
$R^{\rm heat}_{\rm peak}$ & \makecell{Size at which the PDF of the size \\ distribution of the heated regions peaks}    &  Non-uniform $\TS$ model      & 	\\
$R^{\rm HII}_{\rm FWHM}$ &\makecell{ FWHM of the PDF of the size \\ distribution of the \HII regions}     &  Uniform $\TS$ model      & 	\\
$R^{\rm heat}_{\rm FWHM}$ & \makecell{FWHM of the PDF of the size \\ distribution of the heated regions}     &  Non-uniform $\TS$ model      & 	\\
\hline
\hline
\end{tabular}
\label{tab_igm_param}
\end{table*}

\subsection{Derived IGM parameters}
\label{sec_igm_param}
As mentioned above, although {\sc grizzly} uses astrophysical source parameters to  generate brightness temperature maps, the main goal of this work is to infer the IGM properties at $z \approx 9.1$ from the new LOFAR upper limit. At this epoch, the IGM  is expected to consist of \HII regions embedded in a (partially) neutral medium. The signal from such gas is in emission ($\TB > 0$), in absorption ($\TB < 0$) or zero depending on its spin temperature $\TS$. In addition to $\delta_{\rm B}$, two major sources of the spatial fluctuations of the signal are fluctuations in ionized fraction $\XHII$ and spin temperature $\TS$.

\begin{figure}
\begin{center}
\includegraphics[scale=0.55]{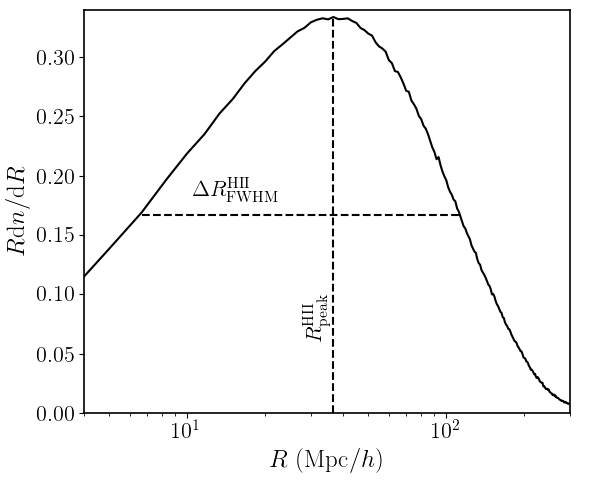}
    \caption{ The probability distribution function (PDF) of the ionized regions of size $R$ estimated using the mean free path method. This ionization state is the same as shown in the left panel of Figure \ref{image_pscool} which corresponds to the parameter choice $\zeta=50$, $M_{\rm min}=3\times 10^{10} ~\MSUN$. $R^{\rm HII}_{\rm peak}$ and $\Delta R^{\rm HII}_{\rm FWHM}$ represent the size of the \HII ~bubbles at which the PDF becomes maximum and the full width at the half maximum of the distribution, respectively. }
   \label{image_bsd}
\end{center}
\end{figure}

If the signal is dominated by $\XHII$ fluctuations, the maximum power spectrum obtained from a model depends not only on the volume averaged ionized fraction ($\overline{\XHII}$) and spin temperature, but also on the size distribution of the \HII bubbles \citep[e.g.][]{Furlanetto2006CharacteristicReionization}. We will therefore study the latter by characterizing the probability distribution function (PDF) of the sizes of \HII~regions with $R^{\rm HII}_{\rm peak}$ and $\Delta R^{\rm HII}_{\rm FWHM}$, which represent the size at which the PDF has a maximum and the full width at half maximum (FWHM), respectively. Figure \ref{image_bsd} shows an example of such a distribution.

Similarly, in the presence of spin temperature fluctuations, the power spectrum of the 21-cm signal also depends on the size distribution of the heated regions (i.e. regions with $\TK>\TCMB$) besides the average gas temperature $\overline{\TK}$, fraction of volume occupied by the heated regions $f_{\rm heat}$ and mass averaged brightness temperature ($\overline{\TB}$). Similarly to the PDF of \HII regions, we will characterise the size distribution of the heated regions adopting the parameters $R^{\rm heat}_{\rm peak}$ and $\Delta R^{\rm heat}_{\rm FWHM}$.

There exists no unique way to characterize the size distribution of a complex three-dimensional structure such as the distribution of \HII/heated~regions in the IGM. We refer the reader to \citet{2011MNRAS.413.1353F}, \citet{Lin2016TheReionization} and \citet{Giri2018BubbleTomography} for an overview of the various methods that can be used. In this work we will use a Monte Carlo based approach, namely the mean free path (MFP) method, first proposed by   \citet{2007ApJ...669..663M}. In the MFP algorithm, we randomly select a point inside the region of interest (e.g.\ \HII regions) and shoot a ray in a random direction until it reaches the boundary of the region. The length of the ray is recorded. When this process is repeated numerous times, the PDF of the recorded lengths provides the PDF of the regions of interest. Here we use $10^7$ rays shot on the fly during the {\sc grizzly} simulation.

A list of parameters used to characterise the IGM is given below (also in Table \ref{tab_igm_param}):
\begin{itemize}
    \item Volume averaged ionized fraction ($\overline{\XHII}$).
    \item Volume averaged gas temperature in the partially ionized IGM with $\XHII< 0.5$ ($\overline{\TK}$).
    \item Uniform spin temperature of the IGM  ($\TS$). Note that 1-(${\TCMB}/{\TS}$) form will be used rather than $\TS$. 
    \item Mass averaged differential brightness temperature ($\overline{\TB}$).
    \item Volume fraction of heated regions with $\TK>\TCMB$ ($f_{\rm heat}$).
    \item $R^{\rm HII}_{\rm peak}$ ($R^{\rm heat}_{\rm peak}$): Size of the \HII (heated) regions at which the probability distribution function (PDF) of the sizes peaks.
    \item $\Delta R^{\rm HII}_{\rm FWHM}$ ($\Delta R^{\rm heat}_{\rm FWHM}$): full width at half maximum of the PDF of the sizes of the \HII (heated) regions.
\end{itemize}

Note that we do not model the signal directly in terms of these IGM parameters, these are rather $derived$ quantities from the simulations.  

\subsection{{\sc grizzly} emulator}
\label{sec:emulator}
Although {\sc grizzly} is fast and  efficient, for parameter estimation with a Bayesian inference framework where hundreds of thousands of models may be needed, the use of  {\sc grizzly} can become computationally too expensive. We therefore adopt an alternative approach. First, we emulate the power spectra derived from {\sc grizzly} simulations using the machine learning algorithm known as Gaussian Process Regression    
\citep[GPR;][]{GPR2006}. The power spectrum emulator is used to interpolate within the parameter space and evaluate the power spectrum for parameter values which have not been simulated.
For a description on how to emulate EoR simulations with GPR, we refer the reader to \citet{2017ApJ...848...23K} and \citet{2019MNRAS.483.2907J}. 
We use the GPR module provided in the python package {\sc scikit-learn} \citep{pedregosa2011scikit}. 
We determine the values for the hyper-parameters for GPR using cross validation, a process which prevents over-fitting of the model \citep[e.g.][]{cawley2010over,hastie01}. We have used 10-fold cross validation \citep{kohavi1995study} to construct the emulators.

Given a set of parameters as described in the previous section, we have configured {\sc grizzly} to generate the spherically averaged power spectrum for the $k$-bins of the LOFAR data. However, as we will see later, not all the data points from the upper limit of the power spectrum are useful for this analysis.  We therefore only use power spectrum amplitudes at scales $k\lesssim 0.15 ~h ~\rm Mpc^{-1}$ to build up our emulator, 
more specifically at $k =$ 0.075, 0.1 and 0.13  $h ~\rm Mpc^{-1}$. 
We quantify the accuracy of the emulators with their mean squared error (MSE)\footnote{The MSE of the emulator is defined as \citep[e.g.][]{2019MNRAS.483.2907J},
\begin{eqnarray}
\mathrm{MSE} = \left< \left(\frac{Q_\mathrm{true}-Q_\mathrm{emulated}}{Q_\mathrm{true}}\right)^2\right>, \nonumber
\label{eq:mse_emul}
\end{eqnarray}
where $<>$ represents the mean estimate. The quantities $Q_\mathrm{true}$ and $Q_\mathrm{emulated}$ are true and emulated values respectively.}. In order to test the accuracy we calculate the MSE for the testing set. The testing set is independent of the data set used for emulation. The MSE of the emulators for predicting the 21-cm power spectrum are found to be less than 10 per cent.

We combine this emulator with the MCMC module available in the {\sc emcee} python package \citep{emcee2013paper} to explore the parameter space of different scenarios. As we are interested in the IGM parameters, we also construct emulators for mapping the source parameters to the IGM parametes. The MSE of these emulators are less than 5 per cent.

\begin{figure*}
\begin{center}
\includegraphics[scale=0.53]{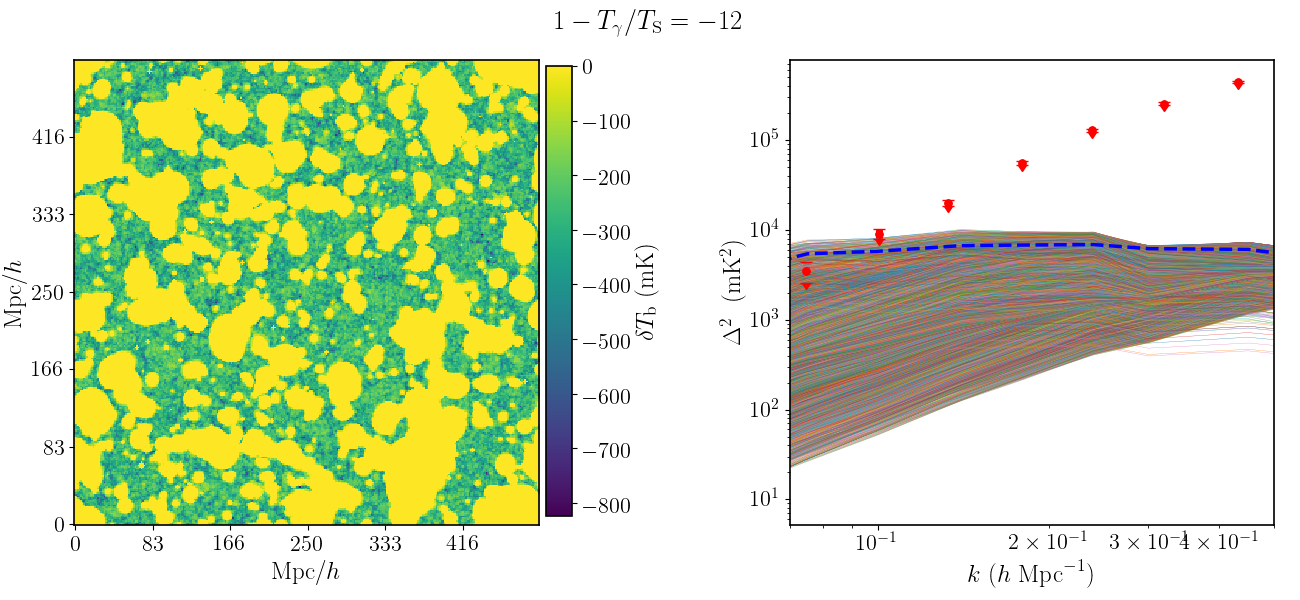}
    \caption{{\it Left panel}: a slice through the brightness temperature cube at $z\approx 9.1$ for the parameter choice $\zeta=50$, $M_{\rm min}=3\times 10^{10} ~\MSUN$ and $1-\TCMB/\TS=-12$. The averaged ionized fraction of this map is 0.55. {\it Right panel}: the curves show the power spectra of the 21-cm brightness temperature as a function of scale for 8556 different combinations of $\zeta$ and $M_{\rm min}$. We assume $1-\TCMB/\TS=-12$, which corresponds to a uniform $\TS=2.1$ K. The red points with error-bars (2-sigma) show the upper limits from the 10-night observations with LOFAR \citep{2019LOFAR}. The dashed blue curve refers to the spherically average power spectrum of the brightness temperature cube from which the slice in the left panel has been extracted. The model power spectra shown in the right panel are also used to build an emulator of the power spectrum using the Gaussian Process Regression.}  
   \label{image_pscool}
\end{center}
\end{figure*}

\begin{figure*}
\begin{center}
\includegraphics[scale=0.53]{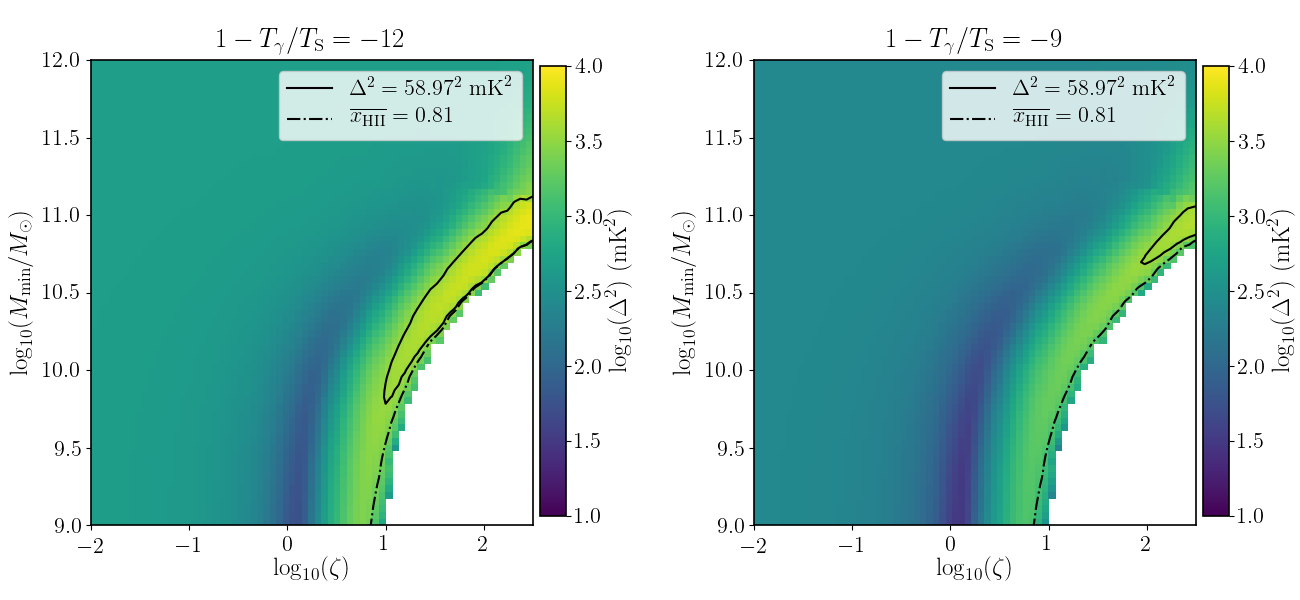}
    \caption{Power spectra at scale $k= 0.075 ~h~\rm Mpc^{-1}$ from the training set as a function of  $\zeta$ and $M_{\rm min}$. We assume  $1-\TCMB/\TS=-12$ and $-9$ for the left and right panels respectively, i.e. $\TS=2.1$ K and 2.73 K at $z \approx 9.1$. The white region at the right bottom corner of the panels corresponds to a fully ionized IGM. The solid contours in both panels represent the upper limit constraint from LOFAR at scale $k=0.075~ h~\rm Mpc^{-1}$, i.e, $\Delta^2 = (58.97)^2$ mK$^2$. For a deterministic observation, the region enclosed by the solid contour will be excluded. The dash-dotted lines denote the contour for $\overline{\XHII}=0.81$.}
   \label{image_pscool2d}
\end{center}
\end{figure*}

\begin{figure*}
\begin{center}
\includegraphics[scale=0.53]{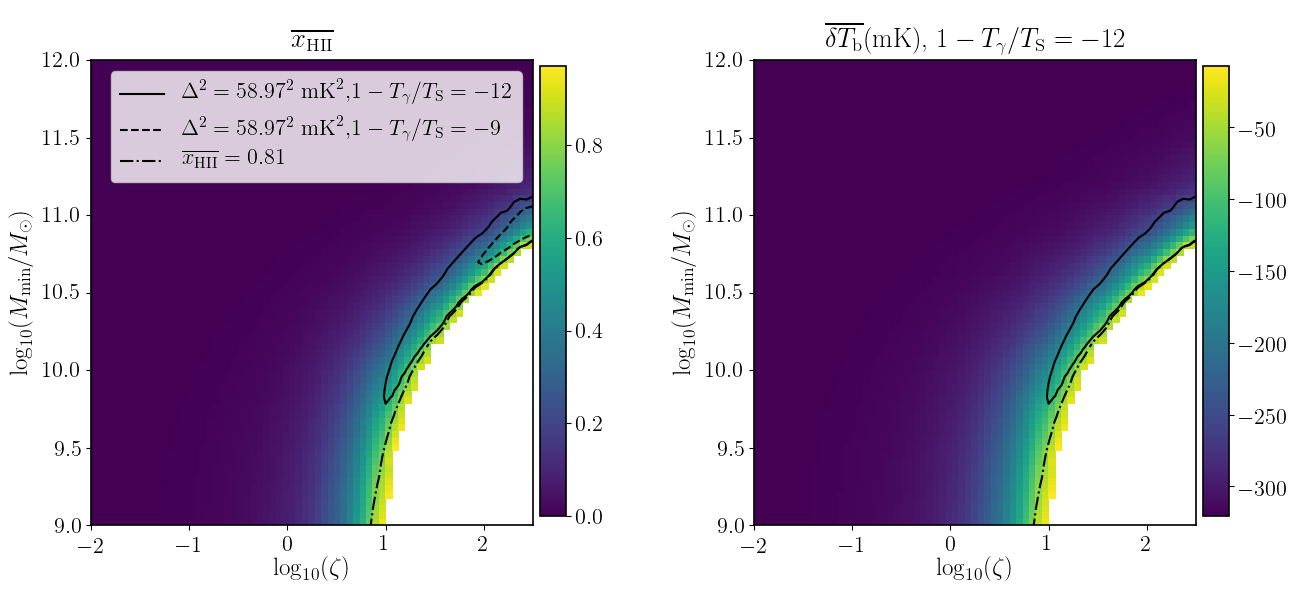}
    \caption{Averaged ionized fraction (left panel) and brightness temperature (right panel) at $z\approx 9.1$ from the training set as a function of  $\zeta$ and $M_{\rm min}$. We assume  $1-\TCMB/\TS=-12$ in the right panel. The white regions at the right bottom corners of the panels corresponds to a fully ionized IGM. The contours in both panels represent the upper limit from LOFAR at scale  $k=0.075 ~h~\rm Mpc^{-1}$, i.e. $\Delta^2=(58.97)^2$~mK$^2$, for $1-\TCMB/\TS=-12$ (solid) and -9 (dashed). For a deterministic observation, the region enclosed by the contours will be excluded. The dash-dotted lines denote the contour $\overline{\XHII}=0.81$.}
   \label{image_globalcool2d}
\end{center}
\end{figure*}

\begin{figure*}
\begin{center}
\includegraphics[scale=0.53]{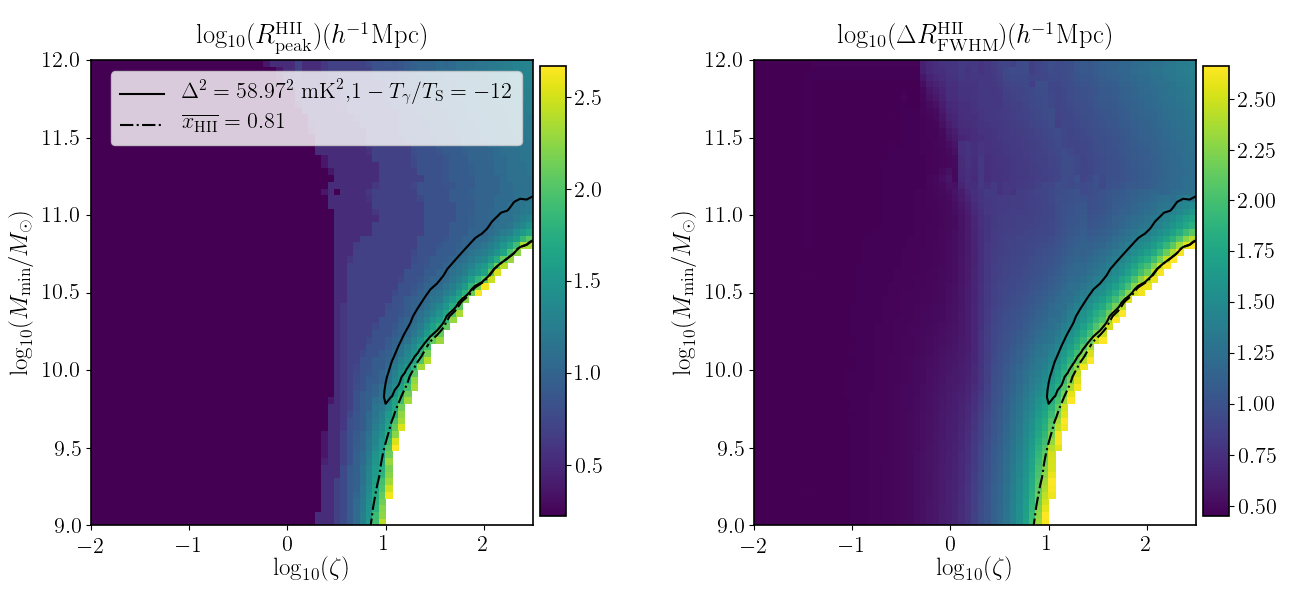}
    \caption{ The distribution of $R^{\rm HII}_{\rm peak}$ (left panel) and $\Delta R^{\rm HII}_{\rm FWHM}$ (right panel) over the parameter space of $\zeta$ and $M_{\rm min}$ for $z\approx 9.1$. $R^{\rm HII}_{\rm peak}$ and $\Delta R^{\rm HII}_{\rm FWHM}$ represent the radius at which the probability distribution of the sizes of the ionized regions has maximum amplitude and the full width at half maximum of that distribution respectively. We use the mean free path method to estimate the bubble size distribution. The region in white in the bottom right of the panels corresponds to fully ionized IGM at redshift $\approx$ 9.1. The solid curves show the contours of $\Delta^2 = (58.97)^2$ mk$^2$ for $1-\TCMB/\TS=$ -12. For a deterministic observation, the region enclosed by the solid contour will be excluded. The dash-dotted line shows the contour for $\overline{\XHII}=0.81$.  }
   \label{image_2dplot_bsd}
\end{center}
\end{figure*}

\begin{figure*}
\begin{center}
\includegraphics[scale=0.7]{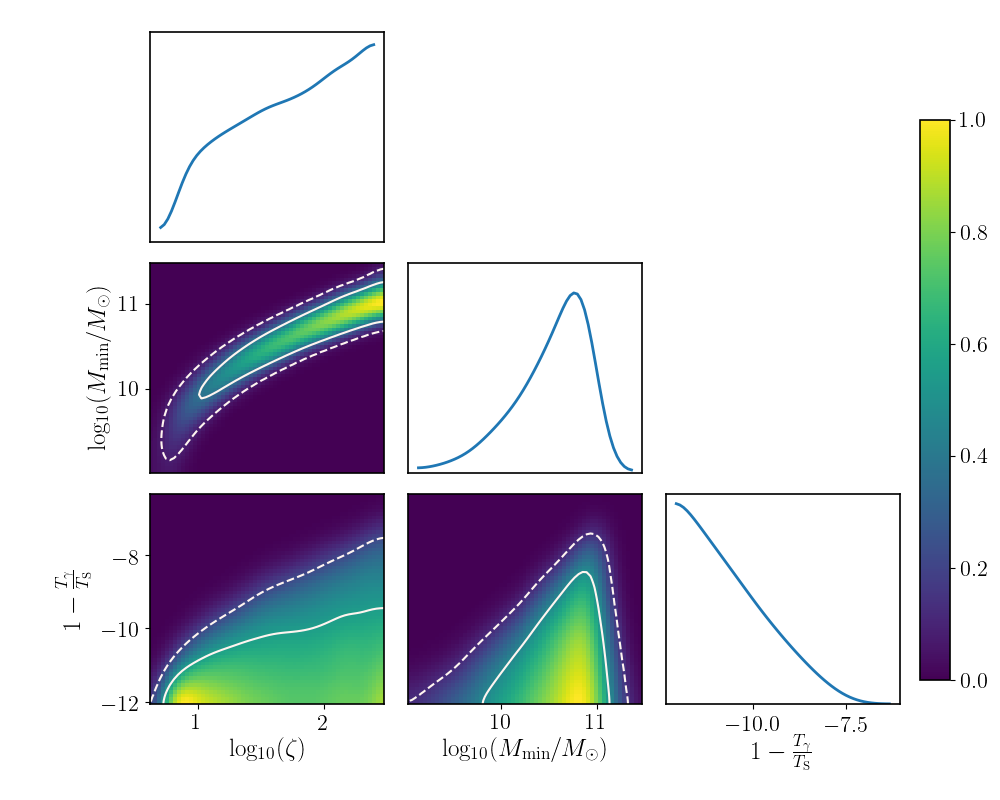}
    \caption{ Constraints on the three {\sc grizzly} parameters of the uniform $\TS$ scenario (see Section~\ref{sec:coldigm}) from the MCMC analysis using the LOFAR upper limit for $z\approx 9.1$. The color-bar shows the probability that models are ruled out. The solid and dashed curves show the 68 and 95 per cent credible intervals of the ruled out models. The diagonal panels show the marginalized probability distribution by which each parameter value as used in the MCMC analysis is ruled out. }
   \label{image_mcmcsambit}
\end{center}
\end{figure*}

\begin{figure*}
\begin{center}
\includegraphics[scale=0.7]{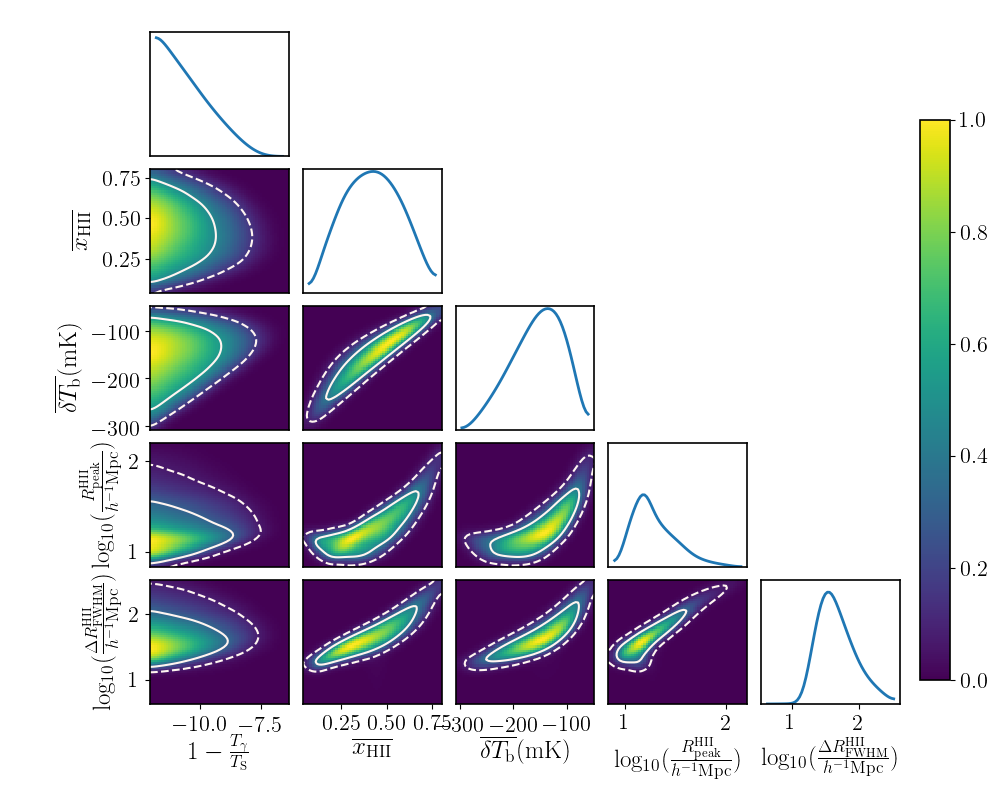}
    \caption{Similar to Figure \ref{image_mcmcsambit}, but this shows constraints on the IGM parameters at $z\approx 9.1$ in the uniform $\TS$ scenario. The color-bar shows the probability that models are ruled out. The solid and dashed curves corresponds to the 68 and 95 per cent credible intervals of the ruled out models. The marginalized probability distributions of the IGM parameters are shown in the diagonal panels. }
   \label{image_igmparam}
\end{center}
\end{figure*}

\subsection{Bayesian inference framework}
\label{sec:like}
As described in the previous section, we combine the {\sc grizzly} emulator with an MCMC algorithm to explore the parameter space for different scenarios and to constrain them using the observed upper limits. The probability of any parameter value $\theta$, i.e. the posterior $p(\mathbf{\theta}|\mathbf{x})$, given some observation $\mathbf{x}$, is defined by Bayes' theorem as,
\begin{eqnarray}
 p(\mathbf{\theta}|\mathbf{x}) \propto p(\mathbf{x}|\mathbf{\theta})~p(\mathbf{\theta}) \,
\end{eqnarray}
where $p(\mathbf{\theta})$ represents the prior on the parameter values. The quantity $p(\mathbf{x}|\mathbf{\theta})$, also known as the likelihood $\mathcal{L}$, gives the probability of any observation given certain parameters.
It should be kept in mind that the likelihood can not be defined by the formal $\chi^2$ method as the observed power spectrum is an upper limit only. Therefore, here we  define the likelihood as follows. %

Let us denote the observed power spectrum $\Delta^2_\mathrm{o} (k_i)$ by $\Delta^2_\mathrm{21} (k_i) \pm \Delta^2_\mathrm{21,err} (k_i)$, while the model power spectrum estimated using the emulator for a set of parameters $\theta$ is denoted by $\Delta^2_\mathrm{m}(k_i, \theta)$. Two major sources of uncertainty on the modelled large-scale power spectrum are: (i) error from the emulators themselves, (ii) sample variance which increases at larger scales. The combined error remains $\lesssim$ 10 per cent for the scales considered in this study. Thus, we assume a 10 per cent modelling error, $\sigma_\mathrm{m} (k_i) = 0.1\times \Delta^2_\mathrm{m}(k_i, \theta)$. 
This error is always larger than the sampling error from the simulation. The total variance $\sigma^2 = \Delta^4_\mathrm{21,err} + \Delta^4_\mathrm{m,err}$ includes the errors from the observation and simulations. For an upper limit, we define the likelihood $\mathcal{L} (\theta)$ for a model with parameters $\theta$ as (see Appendix \ref{appendix:likelihood} for the derivation)
\begin{equation}
    \mathcal{L}(\theta) = \prod_{i} \frac{1}{2}\left[ 1 + \mathrm{erf}\left(\frac{\Delta^2_\mathrm{21} (k_i)- \Delta^2_\mathrm{m}(k_i, \theta)}{\sqrt{2} \sigma (k_i)}\right) \right].
    \label{equ:like}
\end{equation}
This expression results in the following key behaviour: 
\begin{itemize}
    \item  {If the model power spectrum, $\Delta^2_\mathrm{m}(k_i, \theta)$, as estimated using the emulator for a set of parameters $\theta$ is larger than the observed power spectrum, $\Delta^2_\mathrm{21} (k_i)+\sigma (k_i)$, 
    within at least one $k$-bin $k_i$, $\mathcal{L}(\theta)$ approaches 0 and  that model is ruled out by the upper limit.}
    \item  {If $\Delta^2_\mathrm{m}(k_i, \theta)$ is smaller than $\Delta^2_\mathrm{21} (k_i)-\sigma(k_i)$ for {\it all}\/  $k$-bins, $\mathcal{L}(\theta)$ approaches 1, and that model is consistent with the upper limit.} 
    \item  {In case the above two conditions do not hold, the likelihood estimated from Eq. \ref{equ:like} remains between 0 and 1.}
 \end{itemize}
In this work, we aim to find models that are excluded by the measured upper limit. Thus, we use $\mathcal{L}_{\rm ex}(\theta) = 1-\mathcal{L}(\theta)$ in the MCMC analysis as the likelihood of a set of parameters $\theta$ to be excluded by the upper limit. \footnote{Note that, following the same calculation shown in Appendix \ref{appendix:likelihood}, one can also directly estimate the likelihood of set of parameters $\theta$ to be excluded as $\mathcal{L}_{\rm ex}(\theta)= \prod_{i} \frac{1}{2}\left[ 1 - \mathrm{erf}\left(\frac{\Delta^2_\mathrm{21} (k_i)- \Delta^2_\mathrm{m}(k_i, \theta)}{\sqrt{2} \sigma (k_i)}\right) \right].$ }

In addition to this, we use a prior on the ionized fraction estimated from the measured Thomson scattering optical depth in \citet{2018arXiv180706209P}. As we do not have any prior information about the redshift evolution of the average ionized fraction, $\overline{\XHII}$, we estimate the maximum value which is possible at redshift $\approx 9.1$ as follows. If we assume that $\overline{\XHII}$ increases or remains constant  with time, a Thomson scattering optical depth $\tau = 0.054\pm 0.007$ translates into a maximum ionized fraction $0.57\pm 0.24$ at redshift $\approx$ 9.1. Here, we thus use $\overline{\XHII}_{\rm ,max}(z = 9.1)=0.81$ as the maximum possible value for $\overline{\XHII}$ at redshift 9.1.  This corresponds to a scenario in which the universe is neutral at $z>9.1$, has a constant ionized fraction in the range $9.1>z>6$ and reionization ends suddenly at redshift 6.
Note that this is unlikely to be a realistic scenario, as $\overline{\XHII}$ is expected to gradually increase to 1 with time. While studies such as \citet{2015ApJ...809...62P}, \citet{2016MNRAS.455.4295G}, and \citet{2019ApJ...875...67M} consider model dependent reionization histories and estimate $\XHII$ by comparing the estimated $\tau$ for the models with the measured $\tau$ from the CMB observations,  here we use $\overline{\XHII}_{\rm ,max}(z=9.1)=0.81$ as  a model-independent conservative upper limit of the ionized fraction at $z \approx 9.1$.

\section{Results}
\label{sec:results}
Now we apply the parameter estimation framework described in the previous section to the upper limits from \citet{2019LOFAR} (also given in Table \ref{tab_obs}). As described before, we will discuss two scenarios. While the first one considers ionized patches in a uniform $\TS$ IGM, the second one also includes $\TS$ fluctuations. We present our results in the following sections.

\subsection{Ionized patches and a uniform $\TS$}
\label{sec:coldigm}
In this section we focus on the scenario in which the large-scale modes are caused by the presence of ionized regions, and assume a uniform spin temperature with a value $\TS$ (see e.g. \citealt{2015ApJ...809...61A} and \citealt{2015ApJ...809...62P} for previous papers adopting a uniform $\TS$ model). These ionized regions are expected to be photo-heated to a temperature $\TK \approx 10^4$ K and emit no signal as $\XHI \approx 0$. Here, $\TS$ represents the spin temperature of the neutral part of hydrogen in the IGM.  The sizes and spatial distribution of the ionized regions are determined by the astrophysical parameters $\zeta$ and $M_{\rm min}$. Therefore this model has three parameters $\zeta, M_{\rm min}$ and $1-\TCMB/\TS$ which we will explore. 

We further assume the existence of a uniform $\lya$ background which fully couples $\TS$ to the kinetic temperature $\TK$, and thus a uniform $\TS$ implies a uniform $\TK$ for the neutral IGM. The lowest value of $\TK$ is obtained in the complete absence of heating processes, when adiabatic cooling due to cosmological expansion gives $\TK=2.1$~K at $z \approx 9.1$ for our choice of cosmological parameters (calculated using CMBFAST; \citealt{2000ApJS..129..431Z})\footnote{Here we do not consider any additional cooling mechanisms, such as the interaction between baryons and cold dark matter particles, which have been proposed to explain the recent EDGES low-band observations of the global signal at  $z\approx 17$ \citep{EDGES2018, 2018Natur.555...71B} nor additional sources of excess radio background as considered in studies such as \citet{2018ApJ...858L..17F} and  \citet{2019MNRAS.486.1763F} .}.
Higher values for $\TK$ can be caused by heating through X-rays. To obtain a uniform distribution, this heating will have to be driven by very hard rather than soft X-ray photons \citep[see e.g.,][]{pacucci2014, Fialkov14}. 
Since the spin temperature appears in the differential brightness temperature expression (Eq.~\ref{eq:brightnessT}) as $1-\TCMB/\TS$, we use this, rather than $\TS$, as a parameter in our study. For $\TS \gg \TCMB$, $1-\TCMB/\TS$ approaches 1, while for the lowest value of $\TS=2.1$~K, $1-\TCMB/\TS \approx -12$. We therefore explore the range [-12,1].

Since we assume that $1-\TCMB/\TS$ is constant, the power spectrum scales by $(1-\TCMB/\TS)^2$ at all wavenumbers. Therefore, we train our GPR emulator only to generate power spectra for different combinations of  $\zeta$ and $M_{\rm min}$ while keeping $1-\TCMB/\TS=1$. For $\zeta$ and $M_{\rm min}$ we select the ranges [$10^{-2} - 10^{2.5}$] and [$10^9 - 10^{12} ~\MSUN$] respectively. The total number of {\sc grizzly} models used for training the emulator is 8556.

We first illustrate the outcome of this set of {\sc grizzly} models in Figure \ref{image_pscool}. The left panel shows a 2D slice from the brightness temperature cube for the case $\zeta=50$, $M_{\rm min}=3\times 10^{10}~\MSUN$ and $1-\TCMB/\TS=-12$. This combination of parameters produces an ionization map with large \HII bubbles with characteristic size larger than several tens of Mpc and {a volume} averaged ionized fraction of 0.55. The corresponding power spectrum is plotted as a thick dashed curve in the right panel of Figure \ref{image_pscool}, together with the other 8555 models from the training set. All these curves assume the minimal value of $1-\TCMB/\TS=-12$.

The red points in the right panel of Figure \ref{image_pscool} denote the current LOFAR upper limits on $\Delta^2$ with $1-\sigma$ error bars. Clearly, some of the models have a power spectrum amplitude larger than the upper limits at the larger scales. 
These results also show that scales with $k\gtrsim 0.15 ~h ~\rm Mpc^{-1}$ do not significantly constrain the models, which is why, as mentioned in Sect.~\ref{sec:emulator}, we only use the lowest three  $k$ values to build the emulator and to calculate the likelihood in the MCMC framework.

\begin{table*}
\centering
\caption[]{Constraints from the MCMC analysis on the IGM parameters of the uniform $\TS$ scenario at $z\approx 9.1$. Note that our analysis excludes the parameter space that {\it satisfies all the conditions given in this table.}}
\small
\tabcolsep 8pt
\renewcommand\arraystretch{1.5}
   \begin{tabular}{c c c c}
\hline
\hline
\makecell{IGM Parameters of \\ uniform $\TS$ scenario} & Prior & \makecell{68$\%$ credible interval\\ of the excluded models} & \makecell{95$\%$ credible interval\\ of the excluded models}\\
\hline
\hline
$\overline{\XHII}$ & Flat in [0, 0.81]   & $[0.24, 0.60]$   &   $[0.13, 0.74]$ 	\\
$1-\TCMB/\TS$ & Flat in [-12, 1]   &  $[-\infty, -10.21]$  &  $[-\infty, -8.50]$ 	\\
$\TS$ (K) & Flat in [2.1, $\infty$] & [0, 2.435] & [0, 2.874] \\
$\overline{\TB}$ (mK) & --    & $[-189.31, -87.65]$   &   $[-251.23, -56,75]$	\\
$R^{\rm HII}_{\rm peak}$ ($~h^{-1} \rm Mpc$) & --  & $[9.89, 24.55] $   &  $[7.55, 58.07]$ 	\\
$\Delta R^{\rm HII}_{\rm FWHM}$ ($~h^{-1} \rm Mpc$) & -- & $[21.88, 70.79]$   &  $[16.37, 184.93]$ 	\\
\hline
\hline
\end{tabular}
\label{tab_mcmc_igm_units}
\end{table*}

\subsubsection{{\sc grizzly} and IGM parameters}
\label{sec:igmparam_uniTS}
Figure \ref{image_pscool2d} shows the dependence of the power spectra at scale $k= 0.075 ~h~\rm Mpc^{-1}$ on the parameters $\zeta$ and $M_{\rm min}$ obtained from the training set. The left and right panels of the figure correspond to $1-\TCMB/\TS=-12$ and -9 respectively. The solid curves in both panels represent the contours corresponding to the upper limit at this scale, i.e. $\Delta^2 = (58.97)^2~ \rm mK^2$. One can easily see that a significant part of the parameter space can be ruled out by this upper limit alone for $1-\TCMB/\TS=-12$. However, the volume of parameter space which can be excluded rapidly shrinks for higher values of $1-\TCMB/\TS$, and almost no constraints can be set for $1-\TCMB/\TS\gtrsim -8$. 

Figure \ref{image_pscool2d} also shows that the section of parameter space covering $\zeta \gtrsim 10$ and $10^{9.8} ~\MSUN \lesssim M_{\rm min} \lesssim 10^{11} ~\MSUN$ which produces a highly ionized IGM is disfavoured.
In fact, the excluded parameter space remains close to the parameter space which completes reionization by redshift 9.1, which corresponds to the region in white at the bottom right corner of both panels.

In the left panel of Figure \ref{image_globalcool2d}, we plot the average ionized fraction $\overline{\XHII}$ as a function of the  $\zeta$ and $M_{\rm min}$ values we have explored. One can easily see that the models with the largest amplitude of the large-scale power spectrum correspond to an ionized fraction $\approx 0.5$. This is expected as at this stage of the reionization process the typical dimension of the bubbles becomes comparable to the size of the scale of interest \citep[see e.g.,][]{ghara15a}.
As the ionized fraction approaches 1, the power spectrum decreases and becomes negligible at the end of the reionization process due to the paucity of neutral hydrogen. One can see that for $1-\TCMB/\TS=-12$ the excluded parameter space corresponds to average ionized fractions $\gtrsim 0.2$. It is interesting to note that, coincidentally, in this scenario the parameter space excluded by the LOFAR upper limit shares the same boundary at $\overline{\XHII} \approx 0.81$ with the parameter space excluded by the CMB Thomson scattering optical depth constraint on the maximum possible value of ionized fraction at redshift 9.1  (dashed-dotted line, see Section~\ref{sec:like}).

The right panel of Figure \ref{image_globalcool2d} shows the value of the average brightness temperature $\overline{\TB}$ as a function of $\zeta$ and $M_{\rm min}$ for $1-\TCMB/\TS=-12$. $\overline{\TB}$ falls between $\approx -300$ mK (fully neutral) and zero (fully ionized). We find that the excluded parameter space is concentrated around  $\overline{\TB}\gtrsim -250$ mK. This is due to the fact that the average ionized fraction remains $\gtrsim 0.2$ for the excluded parameter space for the case of the lowest spin temperature.

\begin{figure*}
\begin{center}
\includegraphics[scale=0.5]{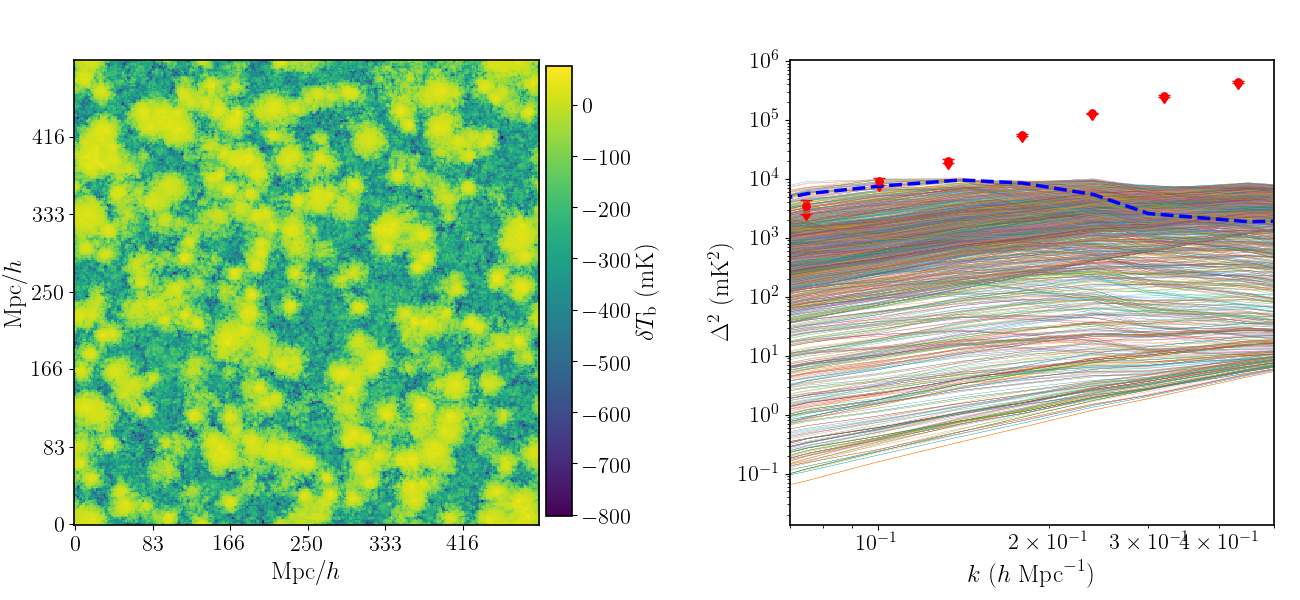}
    \caption{ {\it Left panel}: a slice through the brightness temperature cube for the parameter choice $\zeta=0.1$, $M_{\rm min}=10^{9} ~\MSUN$, $M_{\rm min, X}=3\times 10^{11} ~\MSUN$, $f_X=2$ and $\alpha = 1.2$. The average ionized fraction of this map is 0.01, while the average volume fraction of the heated regions is 0.1. {\it Right panel}: the curves show the power spectra of the 21-cm brightness temperature as a function of scale for 1495 different combinations of parameters. The red points with error-bars show the upper limits from the 10-night observations with LOFAR \citep{2019LOFAR}. The blue dashed curve represents the power spectrum of the brightness temperature cube from which the slice in the left panel has been extracted. } 
   \label{image_psts}
\end{center}
\end{figure*}

\begin{figure}
\begin{center}
\includegraphics[scale=0.45]{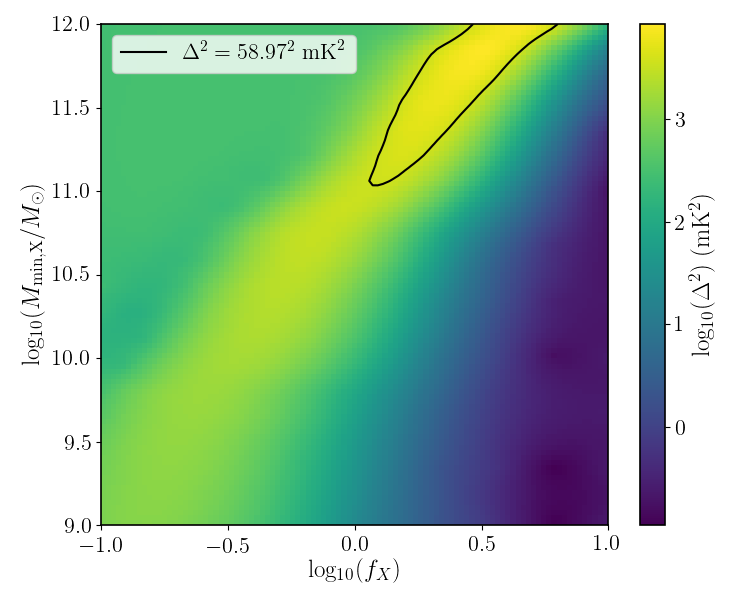}
    \caption{Spherically averaged power spectra of the 21-cm signal from $z\approx 9.1$ at scale $0.075 ~h ~\rm Mpc^{-1}$ as a function of $M_{\rm min, X}$ and $f_X$. These power spectra are generated using an emulator which is trained with 1495 models from {\sc grizzly}. This plot corresponds to $\zeta = 0.1$, $M_{\rm min}=10^9 ~\MSUN$ and $\alpha=1.2$. The solid contour represents the upper limit constraint from LOFAR at scale $0.075 ~h ~\rm Mpc^{-1}$, i.e. $\Delta^2=(58.97)^2$ mK$^2$. For a deterministic observation, the region enclosed by the solid contour will be excluded.    }
   \label{image_2dplot_ps}
\end{center}
\end{figure}

\begin{figure*}
\begin{center}
\includegraphics[scale=0.28]{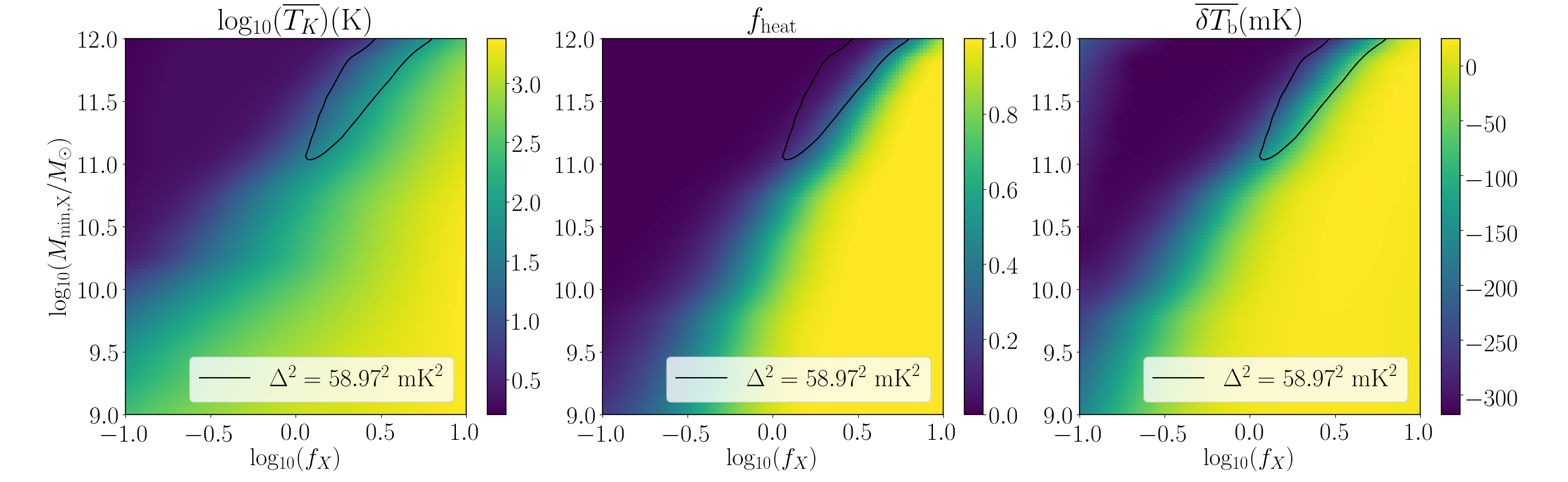}
    \caption{Average gas temperature for regions with ionized fraction less than 0.5 (left panel), volume fraction of the heated regions (middle panel), and  average brightness temperature at redshift 9 as a function of $M_{\rm min, X}$ and $f_X$.  Here $\zeta = 0.1$. The solid curve represents the contour corresponding to $\Delta^2=(58.97)^2$ mK$^2$  at scale $0.075 ~h ~\rm Mpc^{-1}$ which is the LOFAR upper limit on the spherically averaged power spectrum. For a deterministic observation, the region enclosed by the solid contours will be excluded.}
   \label{image_2dplot_tk_femission}
\end{center}
\end{figure*}

\begin{figure*}
\begin{center}
\includegraphics[scale=0.40]{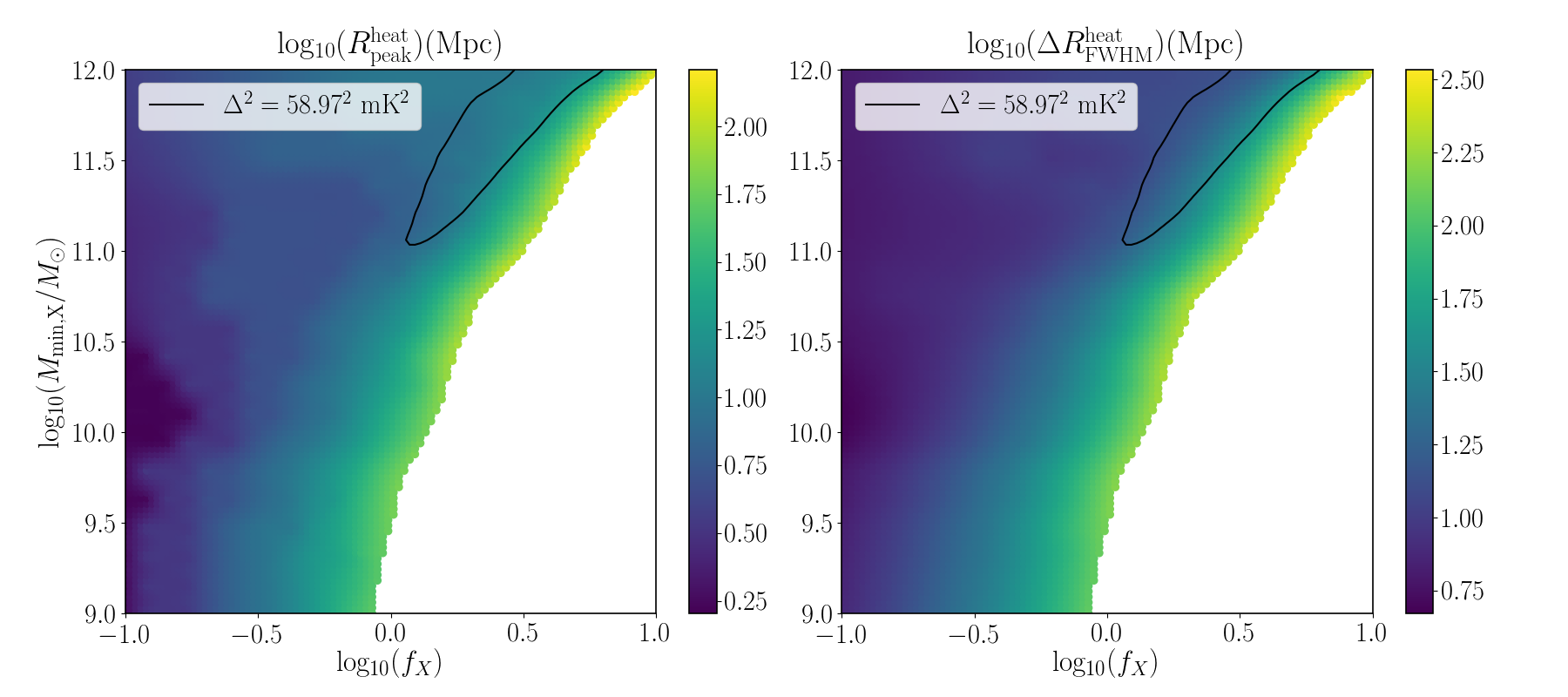}
    \caption{Distribution of $R^{\rm heat}_{\rm peak}$ (left panel) and $\Delta R^{\rm heat}_{\rm FWHM}$ (right panel) at $z\approx 9.1$ as a function of the parameters $f_X$ and $M_{\rm min, X}$ for $\zeta=0.1$. $R^{\rm heat}_{\rm peak}$ and $\Delta R^{\rm heat}_{\rm FWHM}$ represent the radius at which the probability distribution of the sizes of the heated regions (i.e, regions with $\TK>\TCMB$) has maximum amplitude and the full width at half maximum of that distribution, respectively.  We use the mean free path method to estimate the size distribution of the heated regions. The region in white in the bottom right of the panels corresponds to IGM fully heated above the CMB temperature.  The solid curve represents the contour corresponding to $\Delta^2=(58.97)^2$ mK$^2$ which is the LOFAR upper limit on the power spectrum at scale $0.075 ~h ~\rm Mpc^{-1}$. For a deterministic observation, the region enclosed by the solid contours will be excluded.}
   \label{image_2dplot_bsdheat}
\end{center}
\end{figure*}

\begin{figure*}
\begin{center}
\includegraphics[scale=0.7]{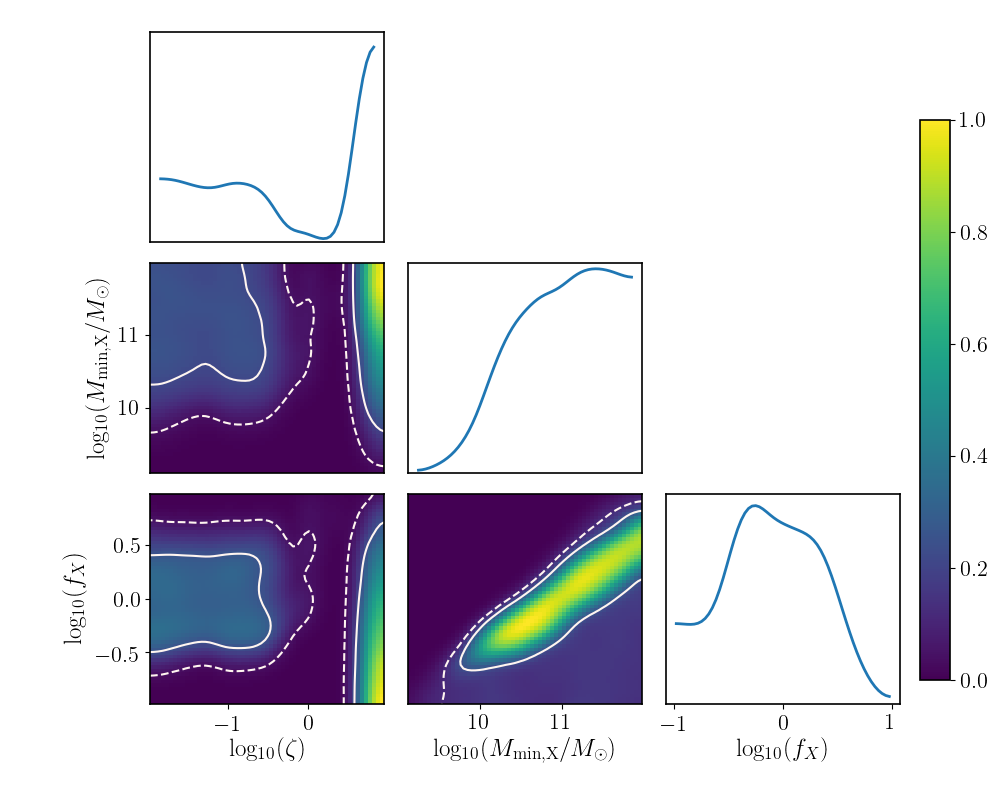}
    \caption{Constraints on the three parameters of our second scenario with non-uniform $\TS$ fluctuations models presented in section~\ref{sec:heat} from the MCMC analysis using the LOFAR upper limit at $z\approx 9.1$. The color-bar shows the probability that models are ruled out. The solid and dashed curves show the 68 and 95 per cent credible intervals of the ruled out models. The diagonal panels show the marginalized probability distribution for the parameters used in the MCMC analysis in this scenario.}
   \label{image_mcmall}
\end{center}
\end{figure*}

\begin{figure*}
\begin{center}
\includegraphics[scale=0.7]{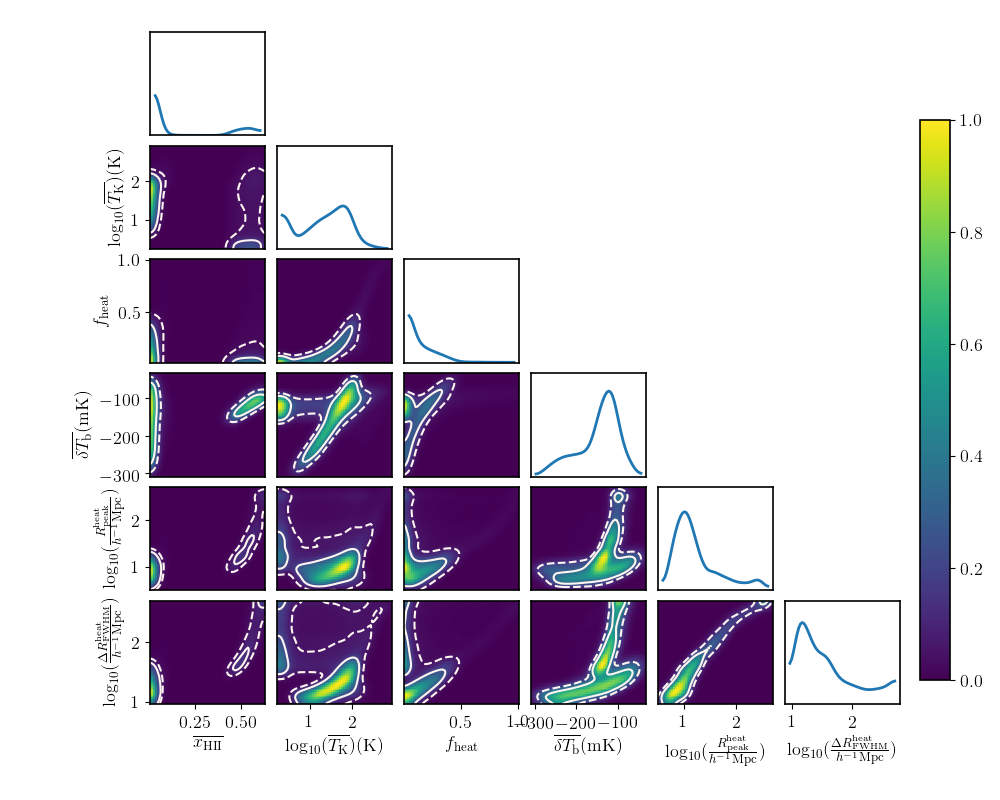}
    \caption{Constraints on the IGM parameters of the non-uniform $\TS$ model presented in section~\ref{sec:heat} from the MCMC analysis using the LOFAR upper limit at $z\approx 9.1$. The color-bar shows the probability that models are ruled out. The solid and dashed curves show the 68 and 95 per cent credible intervals of the ruled out models. The diagonal panels show the marginalized probability distribution for each of the IGM parameters considered in this scenario. } 
   \label{image_mcmalligmhot}
\end{center}
\end{figure*}

Figure \ref{image_2dplot_bsd} shows the dependencies of $R^{\rm HII}_{\rm peak}$ and $\Delta R^{\rm HII}_{\rm FWHM}$ on $\zeta$ and $M_{\rm min}$. Clearly, the most probable size of the bubbles and the FWHM increase with increasing $\zeta$ and decreasing $M_{\rm min}$, as the average ionized fraction increases (also see \citealt{Giri2018BubbleTomography, Giri2018OptimalObservations}).  As expected, the parameter space which is excluded has preferentially a large characteristic size of the ionized regions. More specifically, the part of the parameter space which can be ruled out for $1-\TCMB/\TS=-12$ shows $R^{\rm HII}_{\rm peak}  \gtrsim 10 ~h^{-1} ~\rm Mpc $ and $\Delta R^{\rm HII}_{\rm FWHM}  \gtrsim 30 ~ h^{-1} ~\rm Mpc $. 

To test whether the results for the derived IGM quantities could be sensitive to the choices for the source model, we also explored a source model in which the ionizing emissivity depends non-linearly on the halo mass. As can be seen in Appendix~\ref{appendix:igm}, changing the source model affects the constraints on the source parameters but reproduces the same constraints on the derived IGM parameters, illustrating that these constitute the more robust results of our study.

\subsubsection{MCMC results}
\label{sec:mcmc_units}
Up to this point we have only explored the implications of the LOFAR upper limits using the results from {\sc grizzly} for slices through selected parameter spaces. In this section we employ our parameter estimation framework which includes the emulator results and an MCMC algorithm. The aim is to explore the full parameter space and find the probability that the models are ruled out by the current upper limit from LOFAR. We use 20 walkers and $10^6$ steps for this MCMC analysis. We checked the convergences of the MCMC chains using the integrated auto-correlation time as suggested in \citet{2010CAMCS...5...65G} and find that the chains are well converged for this number of steps. 

The likelihood used for this MCMC analysis is given by Eq \ref{equ:like}. In addition, we use a flat prior on $\overline{\XHII} (z=9.1)\leq 0.81$. Figure \ref{image_mcmcsambit} shows the posterior distribution of the parameters $\zeta$, $M_{\rm min}$ and $1-\TCMB/\TS$, with the solid and dashed curves indicating the 68 per cent and 95 per cent credible intervals\footnote{We estimate the credible intervals of our posterior distributions by the approach based on computing the highest density interval \citep[see e.g.][]{hyndman1996computing}.} of the excluded models within the range of parameters considered here. As expected, and as already suggested by Figure \ref{image_pscool2d}, higher values of $\zeta$ ($\gtrsim 10$) and lower values of $M_{\rm min}$ (in particular $10^{9.8} ~\MSUN \lesssim M_{\rm min}\lesssim 10^{11} ~\MSUN$) are more likely to be excluded as they result in higher ionization and thus a large-scale power spectrum more likely to exceed the observed one. Similarly, a colder IGM is more likely to be ruled out than a hotter IGM, as the former increases the signal strength.

We use a separate emulator to estimate the IGM parameters $\overline{\XHII}$, $R^{\rm HII}_{\rm peak}$ and $\Delta R^{\rm HII}_{\rm FWHM}$ for this scenario from the same set of {\sc grizzly} source parameters as used in our MCMC framework. 
This emulator is constructed using the same method as described in Section~\ref{sec:emulator}. 
Figure \ref{image_igmparam} shows the posterior distribution of the IGM parameters. The constraints on the excluded IGM parameters are also listed in Table \ref{tab_mcmc_igm_units}. 
Clearly, an IGM with $\overline{\XHII} \approx 0.13-0.74$, $(1-\TCMB/\TS) \lesssim -8.5$, and \HII bubble distribution characterised by $R^{\rm HII}_{\rm peak} \approx 8 - 58 ~h^{-1}~\rm Mpc$ and $\Delta R^{\rm HII}_{\rm FWHM} \approx 16-185 ~h^{-1}~\rm Mpc$ is ruled out within 95 per cent credible intervals. This part of the parameter space corresponds to $-250 ~\rm mK \lesssim \overline{\TB} \lesssim -55 ~\rm mK$. Note that the excluded parameter space requires satisfying all of the above-quoted conditions. These results are in agreement with our findings in section \ref{sec:igmparam_uniTS}.   However, it is also clear from Figures \ref{image_mcmcsambit} and \ref{image_igmparam} that tighter constraints on the power spectrum are required to put any bounds on source and IGM parameters with this analysis if the IGM is not very cold.

\subsection{Spin temperature fluctuations}
\label{sec:heat}
 
In this section we relax the uniform $\TS$ assumption and consider the scenario in which X-ray sources cause partial ionization and heating of the IGM. However, we will not  vary all five {\sc grizzly} parameters $\zeta$, $M_{\rm min}$, $M_{\rm min, X}$, $f_X$ and $\alpha$ (see Section \ref{sec:grizzly}). Instead, we fix the values of $M_{\rm min}$ and $\alpha$ and only retain the remaining three parameters. This choice is motivated by a preliminary study suggesting that the LOFAR upper limits provide very weak constraints on $M_{\rm min}$ and $\alpha$.

We set $M_{\rm min}=10^9 ~\MSUN$, i.e. the lowest dark matter halo mass provided by our $N$-body results (Section~\ref{sec:nbody}). This means that, unlike the previous scenario, all halos contribute to the ionization of the neutral hydrogen in the IGM. All the halos also emit $\lya$ photons, building a strong $\lya$ background. We thus assume that the $\lya$ coupling is saturated (in other word, $\TS = \TK$) in this scenario. The value of the X-ray spectral index $\alpha$ is uncertain and dependent on the properties of the X-ray sources. For X-ray sources such as quasars and mini-quasars the spectrum can be very steep, with $\alpha\gtrsim 1$ \citep{2003AJ....125..433V, 2017MNRAS.467.3590G, 2017A&A...608A..51M}, while for high-mass X-ray binaries the observed spectral index can be as small as $\alpha\approx 0.2$ \citep{2012MNRAS.419.2095M, 2019MNRAS.tmp.1386I}.  In this study, we assume $\alpha=1.2$. Below we discuss the effect of different $\alpha$ on our results. 

The remaining three parameters constitute our parameter space. For $\zeta$ we keep the same range used in the previous scenario, while we vary $M_{\rm min, X}$ between $10^9$ and $10^{12} ~\MSUN$, and $f_X$ between 0.1 and 10. As we will see below, this choice covers the regime that is interesting from the point of view of the current LOFAR upper limits. As the run time of the simulations with spin-temperature fluctuations is much longer than in the previous scenario, we initially cover the parameter space with a coarser grid. We then visually identify the part of the parameter space that provides a large amplitude of the large-scale power spectrum and fine sample only that region to increase the accuracy of the emulator.  We thus end up using only 1495 power spectra generated using {\sc grizzly} to train our GPR emulator for this scenario.

In Figure  \ref{image_psts} we show a slice from the brightness temperature map corresponding to the scenario with $\zeta=0.1$, $M_{\rm min, X}=3\times 10^{11} ~\MSUN$ and $f_X=2$.  The average ionized fraction remains $\approx 0.01$ due to the small value of $\zeta$. The average volume fraction of heated regions of this map is also small ($\approx$ 0.1) as  $M_{\rm min, X}$ is large, and thus only a few of the massive halos contribute to the heating. While in the previous scenario the patchiness of the signal  was due to the ionized regions only, now it is also sourced by the heated regions around the sources.

The thick dashed curve in the right panel of Figure \ref{image_psts} refers to the power spectrum of the $\TB$ map shown in the left panel, together with the 1494 other power spectra used to build the three-parameter emulator of $\Delta^2$ for this scenario. Similar to the previous case, we find that the large-scale power spectra of some of the extreme models are larger than the LOFAR upper limits, which are shown by the red data points and their limits in the right panel of the figure. 

\subsubsection{{\sc grizzly} and IGM parameters}
 Figure \ref{image_2dplot_ps} shows the power spectrum at scale $k= 0.075 ~h~\rm Mpc^{-1}$ in the 2D parameter space of  $f_X$ and $M_{\rm min, X}$.  Note that unlike Figure \ref{image_pscool2d}, where the power spectra were derived from the {\sc grizzly} simulations, here they are evaluated directly with the emulator.  In this plot we fix $\zeta=0.1$, which ensures a small average ionized fraction at $z=9.1$ ($\overline{\XHII} = 0.01$).  Clearly, the power spectrum remains the lowest and constant for a combination of a high value of $f_X$ and a low value of $M_{\rm min, X}$. In this case, heating of the partially ionized gas in the IGM due to X-rays becomes very efficient, raising the gas temperature above the CMB ($\TK \gg \TCMB $) and rendering $\TB$ independent of the spin temperature. On the other hand, the heating of the gas in the IGM remains inefficient for a combination of small $f_X$ and high $M_{\rm min, X}$. As expected, the power spectrum for such models (top left corner of the Figure) is larger than the power spectrum for the heated IGM (bottom right corner of the Figure).  

One can see that the spin temperature fluctuations are efficient around the diagonal of the parameter space, starting from small values of $f_X$ and $M_{\rm min, X}$. Specifically, the combination of high $M_{\rm min, X}$ and $f_X$ enhances the large-scale power spectrum. In this combination, the heated/emission regions around the rare X-ray emitting sources remain isolated in the background absorption signal (e.g., see the left panel of Figure \ref{image_psts}). Also, the partial ionization and heating of the IGM far away from the X-ray emitting sources remain small for a high value of $M_{\rm min, X}$. We also plot the contour corresponding to the LOFAR current upper limit at scale  $k= 0.075 ~h~\rm Mpc^{-1}$, i.e. $(58.97)^2$ mK$^2$. Clearly, some parts of the parameter space with the combination of high $M_{\rm min, X}$ ($\gtrsim 10^{11} ~\MSUN$) and $f_X$ ($\gtrsim 1$) are ruled out with high confidence.

Next, we will consider the global parameters of this scenario. Note that to estimate the IGM parameters we use an emulator different from the one used for the source parameters. In Figure \ref{image_2dplot_tk_femission} we show the average temperature  ($\overline{\TK}$) of regions with ionized fraction smaller than 0.5, the volume fraction of heated regions $f_{\rm heat}$, and the average brightness temperature $\overline{\TB}$.  As expected, $\overline{\TK}$ remains small for a combination of high $M_{\rm min, X}$ and low $f_X$, which also keeps $f_{\rm heat}$ low. In this case the average signal remains in absorption, as shown in the right panel of the figure.  On the other hand, $\overline{\TK}$ is high for the opposite case of a low $M_{\rm min, X}$ and a high $f_X$, for which $f_{\rm heat}$ approaches 1 and $\overline{\TB}$ becomes positive. The parameter space excluded by the LOFAR upper limit at scale $k= 0.075 ~h~\rm Mpc^{-1}$ is shown by the solid curves in all panels. It corresponds to 10~K $ \lesssim \overline{\TK} \lesssim 100$ K, $f_{\rm heat} \lesssim 0.3$ and -200~mK $ \lesssim \overline{\TB} \lesssim -100$ mK. 

In this scenario, the size distribution of the heated regions is more relevant than the size distribution of the ionized regions.
Similarly to the size distribution of the ionized regions considered in the previous scenario, here we analyze the size distribution of the heated regions, characterising the PDF with two parameters, namely $R^{\rm heat}_{\rm peak}$ and $\Delta R^{\rm heat}_{\rm FWHM}$, which represent the size of the heated regions at which the PDF becomes maximum and full width of the half maximum of the PDF, respectively.
Figure \ref{image_2dplot_bsdheat} shows the distribution of $R^{\rm heat}_{\rm peak}$ and $\Delta R^{\rm heat}_{\rm FWHM}$, suggesting that the characteristic size of the heated regions increases with increasing $f_X$. The white regions represent an IGM fully heated above the CMB temperature. The parameter space in the range  $R^{\rm heat}_{\rm peak} \approx 5-20 ~h^{-1}$ Mpc and $\Delta R^{\rm heat}_{\rm FWHM} \approx 10-30 ~h^{-1}$ Mpc  is the one excluded by the LOFAR upper limit at $z=9.1$. Note that the excluded parameter space requires satisfying all of the above-mentioned conditions.

\subsubsection{MCMC results}
Next, we explore the three-dimensional parameter space, i.e. $\zeta$, $M_{\rm min, X}$ and $f_X$, using MCMC to find models that are ruled out by the current LOFAR upper limit. Similar to our previous scenario, we have used 20 walkers and $10^6$ steps for the MCMC analysis and checked the convergence of the chains. Note that we use a flat prior on $\overline{\XHII} (z=9.1)\leq 0.81$. The outcome of the analysis is summarized in Figure \ref{image_mcmall}.  Clearly, a high emissivity of X-ray photons ($f_X \gtrsim 0.3$) with a large $M_{\rm min, X}$ ($\gtrsim 10^{10} ~\MSUN$) is the most likely to be excluded within the 68 per cent credible intervals by LOFAR alone. This combination of parameter values results in large heated regions around rare massive halos embedded in a cold IGM. On the other hand, the combination of large $f_X$ and a small $M_{\rm min, X}$ causes more uniform heating and thus it reaches more easily the  $\TS \gg \TCMB$ condition where the power spectrum remains lower than the measured one.  
Similarly, a very small value of $f_X$ yields almost no heating and coincides with the scenario discussed in the previous section. In such models, a larger value of $\zeta$ is more likely to be ruled out as we have also seen in the previous scenario. Therefore we see a second ruled out region in the parameter space shown in Figure \ref{image_mcmall}.

Next, we will constrain the IGM parameters of this non-uniform $\TS$ scenario, and show the posterior distribution of the IGM parameters in Figure \ref{image_mcmalligmhot}. These results are also listed in Table \ref{tab_mcmc_igm_nonunits}. Clearly, two regimes of the parameter space are likely to be excluded. The first one has large \HII regions in a poorly heated IGM, which is the configuration already discussed in the previous section. In this case, least likely values of the IGM parameters are: $0.5\lesssim \overline{\XHII} \lesssim 0.6$, $\overline{\TK} \lesssim 3.55$ K with  $f_{\rm heat}\approx 0$. 
The second part of the parameter space which is likely to be excluded corresponds to large heated regions with: $\overline{\XHII}\lesssim 0.08$, 7~K $\lesssim\overline{\TK} \lesssim 160$ K,  $-234 \lesssim \overline{\TB}\lesssim -65 $ mK, $f_{\rm heat} \lesssim 0.34$, $3.5 ~h^{-1}{\rm Mpc}\lesssim R^{\rm heat}_{\rm peak}\lesssim 70 ~h^{-1}$ Mpc and $ \Delta R^{\rm heat}_{\rm FWHM}\lesssim 110~h^{-1}$ Mpc. These limits correspond to 95 per cent credible intervals as shown in Figure \ref{image_mcmalligmhot}.

Up to this point we have only considered $\alpha=1.2$. A less steep SED with a smaller value of $\alpha$ contains a smaller number of soft X-ray photons and a larger number of hard X-ray photons. Thus, the heating due to an X-ray spectrum with smaller $\alpha$ is less patchy than that from a steeper spectrum \citep[see e.g.,][]{ pacucci2014, 2017MNRAS.469.1166D, 2019MNRAS.tmp.1386I}, resulting in a smaller amplitude of the large-scale power spectrum of the signal. We have verified that for $\alpha=0.3$ the results are similar to those obtained with $\alpha=1.2$, except that the contour of the excluded region (see Figure \ref{image_2dplot_ps}) shrinks towards higher $M_{\rm min, X}$ values and it shifts slightly towards higher values of $f_X$.

\begin{table*}
\centering
\caption[]{Constraints from the MCMC analysis on the IGM parameters of the non-uniform $\TS$ scenario at $z\approx 9.1$. Note that our analysis excludes the parameter space that satisfies all the conditions given in this table.}
\small
\tabcolsep 12pt
\renewcommand\arraystretch{1.5}
   \begin{tabular}{c c c c c c}
\hline
\hline

\makecell{IGM Parameters of \\ non-uniform $\TS$ scenario} & Prior & \makecell{68$\%$ credible interval\\ of the excluded models } & \makecell{95$\%$ credible interval\\ of the excluded models }  \\

\hline
\hline
$\overline{\XHII}$ & Flat in [0, 0.81]   &  [0, 0.06], [0.50, 0.58]  &  [0, 0.08], [0.45, 0.62] &  	\\
$\overline{\TK}$ (K) & Flat in $[2.10,\infty)$     & $[19.23, 115.61]$, [$2.10, 2.32$]   &    $[7.41, 158.48]$, $[2.10, 3.55]$ &	\\
$f_{\rm heat}$ & -- & [$0, 0.14$]  &  [$0, 0.34$] & \\
$\overline{\TB}$ (mK) & --   & $[-154.50, -84.26]$   & $[-234.15, -65.53]$      & 	\\
$R^{\rm heat}_{\rm peak}$ ($~h^{-1} \rm Mpc$) & --  & $[5.32, 17.78]$   &  $[3.50, 69.82]$  &	\\
$\Delta R^{\rm heat}_{\rm FWHM}$ ($~h^{-1} \rm Mpc$)& -- &  $[10.47, 38.01]$  &  $[0, 113.76]$    & 	\\
\hline
\hline
\end{tabular}
\label{tab_mcmc_igm_nonunits}
\end{table*}

\section{Discussion}
\label{sec:discussion}
We have considered two extreme scenarios, one in which fluctuations at large scales are driven by large ionized regions in a uniform spin temperature IGM, and the other in which they are driven by large heated regions in a non-uniform spin temperature IGM. 
One question that naturally arises is whether there exist other models capable of exceeding the LOFAR upper limits which are not covered by the two scenarios we have explored. As fluctuations in the 21-cm signal are induced by ionization and/or spin temperature fluctuations, it seems hard to come up with alternative scenarios which can be excluded without invoking non-standard physics.

A second question is whether the extreme cases considered are in any way realistic or whether they are already excluded by other observations. We have limited ourselves to deriving constraints from the LOFAR upper limits at $z=9.1$ and have not added information from other redshifts, apart from a very conservative upper limit on the ionized fraction based on the Thomson scattering optical depth derived from the Planck results. This has been a conscious choice as using data from multiple redshifts requires assumptions about the evolution of the source properties which, given the small constraining power of the LOFAR upper limits, does not seem justified. However, it is still possible to apply a minimal check on the models that we find to be excluded by the LOFAR upper limits.

We first consider the scenario in which the excluded models require a fairly large value for $\overline{\XHII}$. The results from \citet{Mitra15} show that the combined constraints from Planck and $z>6$ quasar spectra imply that $\overline{\XHII} \lesssim 0.4$ at $z=9.1$. Although this limits the constraining power of the LOFAR upper limits, the latter is still unique in excluding some models, as we have found cases with $\overline{\XHII}\approx 0.3$ and $1-\TCMB/\TS=-12$ which violate them (see Figure \ref{image_globalcool2d}).

\citet{monsalve2017} presented phenomenological constraints on the evolution of the global 21-cm signal derived from EDGES high-band observations. These constraints are mostly about changes in the signal and are therefore very different from the single $z=9.1$ upper limits used for our results. However, our approach does produce values for the global signal (see the right hand panels in Figures~\ref{image_globalcool2d} and ~\ref{image_2dplot_tk_femission}), with excluded models lying in the range -250~K $ \lesssim \overline{\TB} \lesssim -55$ mK. These can be compared to the values in Figure~9 of \citet{monsalve2017}, where the authors show that for a minimum value of -200~mK, the $\Delta z$ for the full width half maximum of the entire absorption feature has to be above $\approx 5$. They also show that this lower limit is inconsistent with an end of reionization at $z\approx 6$. At face value this implies that the models excluded by the new LOFAR upper limits on the 21-cm power spectrum are also excluded by the EDGES high-band constraints on the evolution of the global signal. However, it should be kept in mind that the EDGES constraints are based on an assumed Gaussian profile for the evolution of the global signal. Furthermore, the systematics for the EDGES results are not fully known \citep[e.g.][] {2018Natur.564E..32H, 2019ApJ...880...26S}.

This comparison to previous results shows that the new LOFAR upper limits exclude rather extreme models which were already unlikely in view of other observational constraints. However, it is important to point out that the LOFAR observations are of a very different character and thus contribute a new and independently obtained piece of the reionization puzzle. As we obtain more stringent upper limits and additional redshift points, the constraints will improve and start to rule out increasingly large regions of the parameter space.

\section{Summary \& Conclusions}
\label{sec:con}

In this paper, we have used the new LOFAR upper limit on the dimensionless spherically averaged power spectrum of the 21-cm signal from redshift $\approx 9.1$ \citep{2019LOFAR}  and investigated which reionization scenarios can be ruled out by it. The upper limits as obtained from 10 nights of observations are $(58.97)^2$ mK$^2$ and $(95.21)^2 ~\rm mK^2$ at scales $k=0.075$ and 0.1 $ ~h$ Mpc$^{-1}$,  respectively.  As these numbers are much larger than the amplitude of the power spectrum expected for standard reionization histories, we mainly focused on the extreme models that produce such high values for the large-scale power spectrum. However, our study also covers the usual range of the parameter space.

With the code {\sc grizzly} we generated power spectra for thousands of models for different combinations of parameters namely, ionization efficiency ($\zeta$), minimum mass of the UV emitting halos ($M_{\rm min}$), minimum mass of X-ray emitting halos ($M_{\rm min, X}$) and X-ray heating efficiency ($f_X$).  
On the basis of these results we build emulators for different scenarios based on Gaussian process regression that map source parameters to power spectra.
These emulators combined with an MCMC framework are then used to constrain the source parameters at $z\approx 9.1$ using the observed upper limits. We also build emulators that map source parameters to IGM parameters, which are used to put constraints on the IGM parameters.
We considered two extreme scenarios in which large-scale fluctuations of the signal are driven by (i) ionized regions embedded in an IGM with a uniform spin temperature, (ii) spin temperature fluctuations. 

As the 21-cm observations themselves characterise the state of the IGM, a major focus of this study is to constrain the thermal and ionization state of the IGM at $z\approx 9.1$ using these upper limits. We study the state of the IGM in terms of parameters such as the average ionization fraction ($\overline{\XHII}$), average gas temperature of the partially ionized IGM ($\overline{\TK}$), (1-$\TCMB/\TS$), mass averaged brightness temperature ($\overline{\TB}$), volume fraction of the heated region ($f_{\rm heat}$), size of the \HII(heated) regions at which the PDF of the sizes peaks $R^{\rm HII}_{\rm peak} (R^{\rm heat}_{\rm peak})$ and the FWHM of the PDFs $\Delta R^{\rm HII}_{\rm FWHM} (\Delta R^{\rm heat}_{\rm FWHM})$. 
The results of our study can be summarized as follows.

\begin{itemize}
\item In the uniform $\TS$ scenario, we found that the models which can be ruled out by the upper limit have a high UV photon emission rate. More specifically, the model with the coldest possible IGM, i.e. $\TS \simeq 2.1$ K, requires an emission rate $\gtrsim 2.85\times 10^{46} \rm s^{-1} ~\MSUN^{-1}$, which is 10 times larger than that predicted by population synthesis codes. At the same time, those models require a suppression of ionizing photons from halos with mass $\lesssim 10^{9.8} ~\MSUN$.  

\item A high emissivity of the UV photons renders the gas in the IGM largely ionized at the target redshift, so that ionized fractions $\overline{\XHII}\gtrsim 0.13$ are excluded within a $95$ per cent credible interval. At the same time, the \HII ~bubbles required have to be few in number and large in size.  The characteristic size of the \HII ~bubbles needs to be, $R^{\rm HII}_{\rm peak}\gtrsim 8 ~h^{-1} \rm Mpc $, with a FWHM of the probability distribution of the size distribution larger than $16 ~h^{-1} \rm Mpc $. This keeps the average brightness temperature of the excluded models $\gtrsim -250$ mK.
The size of the parameter space which can be excluded depends crucially on the value of $\TS$, as it decreases with increasing $\TS$ and no constraints can be set for $\TS\gtrsim 3$ K.  

\item For the scenario where the large-scale fluctuations of the signal are driven by spin temperature fluctuations, we found that the models ruled out are those in which regions with temperature larger than CMB cover a volume fraction $\lesssim 0.34$ and at the same time are large with a characteristic size in the range $3.5-70 ~h^{-1} ~\rm Mpc$ and a size distribution with a FWHM of $\lesssim 110 ~h^{-1} ~\rm Mpc$. The average gas temperature of the partially ionized regions for these excluded models is 7-160 K, while the average brightness temperature lies in between -234 mK and -65 mK.  The heated regions required for these excluded models are large in size and few in number at the same time. This implies that scenarios in which the heating is driven by fewer X-ray emitting sources hosted by the rare massive halos ($M_{\rm min,X} \gtrsim 10^{10} ~\MSUN$) with a high emissivity of X-ray photons (X-ray luminosity $\gtrsim 10^{34} ~\rm erg ~s^{-1} ~\MSUN^{-1}$) are more likely to be ruled out by the current upper limit.

\end{itemize}

As the current upper limits on the 21-cm power spectrum are rather large and restricted to one redshift, the constraints on the IGM and source parameters that can be obtained are not yet very strong. However, they do illustrate the potential of this type of observations to characterise the state of the IGM and from this the properties of early sources in a redshift range which has not been yet well explored. We expect LOFAR to produce more stringent upper limits on the power spectrum both through analysing more of the available data (also at other redshifts) and improving the methods to deal with systematic effects. Combining these with other observables, such as the global 21-cm signal and observations of high-z galaxies using the present and next generation of ground based and space telescopes such as the James Webb Space Telescope, the European Extremely Large Telescope and the Atacama Large Millimetre Array, will give us a much deeper understanding of this crucial period in the history of the Universe.

\section*{Acknowledgements}
The authors would like to thank Jens Jasche and Daniel J. Mortlock for useful discussions,  and an anonymous referee for insightful comments. We acknowledge
that the results in this paper have been achieved using the PRACE Research Infrastructure
resources Curie based at the Tr$\grave{\rm e}$s Grand Centre de Calcul (TGCC) operated by CEA near
Paris, France and Marenostrum based in the Barcelona Supercomputing Center, Spain. Time
on these resources was awarded by PRACE under PRACE4LOFAR grants 2012061089 and
2014102339 as well as under the Multi-scale Reionization grants 2014102281 and 2015122822. We have used resources provided by the Swedish National Infrastructure for Computing (SNIC) (proposal number SNIC 2018/3-40) at PDC, Royal Institute of Technology, Stockholm. GM, RG and SKG are thankful for support by Swedish Research Council grant 2016-03581. R.G. and S.Z. furthermore acknowledge support by the Israel Science Foundation (grant no. 255/18). VJ acknowledges support by the Croatian Science Foundation for a project IP-2018-01-2889 (LowFreqCRO). FGM and LVEK acknowledge support from a SKA-NL roadmap grant from the Dutch Ministry of OCW. EC acknowledges support from the Royal Society via. the Dorothy Hodgkin Fellowship.

\bibliography{mybib,Mendeley_SG,manual}

\appendix
\section{GRIZZLY}
\label{app:grizzly}

Here we briefly describe the one-dimensional radiative transfer method used in the code {\sc grizzly} to simulate the redshifted 21-cm signal from the EoR. We refer the reader to \citet{ghara15a, ghara18} for a more detailed description of the method. While the basic approach of {\sc grizzly} mainly follows the {\sc bears} algorithm  \citep{Thom08, Thom09, Thom11}, it  differs slightly from the original method. Both codes avoid solving the one-dimensional radiative transfer equations on the fly and, instead, they use previously generated 1D ionization profiles to simulate an ionization field. Below, we briefly describe the steps used in {\sc grizzly}: 

\begin{itemize}
\item  First, we generate a large number of 1D profiles of ionized fraction and kinetic temperature for different combinations of source parameter values. The parameters used for this are ionization efficiency, the ratio of X-ray and UV luminosities, X-ray spectral index,  over-density of the uniform background IGM and redshift. In this study, we assume that the age of the source is 10 Myr. For a given cosmology, these profiles need to be generated only once. 

\item Next, we determine the size of the \HII ~regions in all the 1D profiles and create a list of their radii for different parameter values. These are defined as the distance from the center of the source at which the ionized fraction drops to 0.5. 

\item Given a halo with a certain mass and position, we first determine the corresponding UV luminosity. From this, we determine the size of the \HII ~region around that halo using the density field and the list of radii as generated in the previous step. This is done iteratively as follows. We start with a small value of the radius, estimate the spherically averaged over-density contained within it and look for the same combination of radius and over-density in the pre-compiled list. If this is not found, we change the initial choice of the radius and continue the iteration until a match is obtained.
This step is repeated for all halos. The corresponding ionization profiles are used to generate the ionization field.

\item When individual \HII ~regions overlap, we estimate the number of photons in excess and distribute them around the surface of the overlapping regions so that all the photons are used for ionization.

\item We then generate the kinetic temperature field from the ionization field and a  correlation of the ionized fraction and the gas temperature \citep[for details, see][]{ghara15a}.

\end{itemize}

Given the value of the uniform $\TS$, the $\TB$ maps can be generated using the ionization maps and density field for our first scenario. For our second scenario, which assumes $\TS=\TK$, we use the ionization, density and temperature maps to generate the $\TB$ maps following Eq. \ref{eq:brightnessT}.

\section{Likelihood for upper limit observations}
\label{appendix:likelihood}
Using Bayes theorem, we can write the posterior of our model parameters $\mathbf{\theta}$ for simulating the model power spectrum $\Delta^2_\mathrm{m}(\mathbf{k}, \theta)$ given the observed power spectrum $\Delta^2_\mathrm{o}(\mathbf{k})$ as follows,
\begin{eqnarray}
 p(\mathbf{\theta}|\Delta^2_\mathrm{o}(\mathbf{k})) \propto p(\Delta^2_\mathrm{o}(\mathbf{k})|\mathbf{\theta}))~p(\mathbf{\theta})
\end{eqnarray}
\noindent where the first and second term in the right hand side of the equation are the likelihood $\mathcal{L}(\mathbf{\theta}|\Delta^2_\mathrm{o}(\mathbf{k}))$ and prior, respectively.

If $\Delta^2_\mathrm{o}(\mathbf{k})$ is a deterministic, a scenario is ruled out when the modelled power spectrum $\Delta^2_\mathrm{m}(k_i)$ is above $\Delta^2_\mathrm{o}(k_i)$ in any one wavenumber-bin $k_i$. We can write $\mathcal{L}(\theta|\Delta^2_\mathrm{o}(k_i))$ as a Heaviside function $\mathcal{H}(\Delta^2_\mathrm{o}(k_i)-\Delta^2_\mathrm{m}(k_i))$. However, the $\Delta^2_\mathrm{o}(k_i)$ is probabilistic with mean of $\Delta^2_\mathrm{21} (k_i)$ and standard deviation of  $\Delta^2_\mathrm{21,err} (k_i)$. Therefore we need to draw a $\Delta^2_\mathrm{a}(k_i)$ from a normal distribution $\mathcal{N}\left(\Delta^2_\mathrm{21} (k_i), \Delta^2_\mathrm{21,err} (k_i)\right)$ 
and calculate the probability of our model. Here $\Delta^2_\mathrm{a}(k_i)$ is a nuisance parameter over which we can marginalise to get the $\mathcal{L}(\theta|\Delta^2_\mathrm{o}(k_i))$. 
Therefore $\mathcal{L}(\theta|\Delta^2_\mathrm{o}(k_i))$ can be written as follows,
\begin{eqnarray}
\label{eq:marginalise_Pa}
\begin{split}
 \mathcal{L}(\theta|&\Delta^2_\mathrm{o}(k_i))  \\
 &= \int^\infty_{-\infty} p(\Delta^2_\mathrm{o}(k_i)|\Delta^2_\mathrm{a})~p(\Delta^2_\mathrm{a}(k_i)|\theta)~\mathrm{d}\Delta^2_\mathrm{a}.
 \end{split}
\end{eqnarray}

The value for $p(\Delta^2_\mathrm{a}(k_i)|\theta)$ is a Heaviside function $\mathcal{H}(\Delta^2_\mathrm{a}(k_i)-\Delta^2_\mathrm{m}(k_i))$, while $p(\Delta^2_\mathrm{o}(k_i)|\Delta^2_\mathrm{a}(k_i))$ is defined by a $\mathcal{N}\left(\Delta^2_\mathrm{21} (k_i), \Delta^2_\mathrm{21,err} (k_i)\right)$.  
Putting these functions into equation~\ref{eq:marginalise_Pa}, we get
\begin{eqnarray}
\label{eq:marginalise_Pa1}
\begin{split}
&\mathcal{L}(\theta|\Delta^2_\mathrm{o}(k_i)) \\
&=\frac{1}{\sqrt{2\pi}\Delta^2_\mathrm{21,err}}\int^\infty_{-\infty} \mathcal{H}(\Delta^2_\mathrm{a}(k_i)-\Delta^2_\mathrm{m}(k_i))~e^{-\frac{1}{2}\left(\frac{\Delta^2_\mathrm{a}(k_i)-\Delta^2_\mathrm{21}(k_i)}{\Delta^2_\mathrm{21,err}}\right)^2}~\mathrm{d}\Delta^2_\mathrm{a} \\
&=\frac{1}{\sqrt{2\pi}\Delta^2_\mathrm{21,err}} \int^\infty_{\Delta^2_\mathrm{m}(k_i)} e^{-\frac{1}{2}\left(\frac{\Delta^2_\mathrm{a}(k_i)-\Delta^2_\mathrm{21}(k_i)}{\Delta^2_\mathrm{21,err}}\right)^2}~\mathrm{d}\Delta^2_\mathrm{a} \\
&= \frac{1}{2}\left[1+\mathrm{erf}\left(\frac{\Delta^2_\mathrm{21} (k_i)-\Delta^2_\mathrm{m}(k_i)}{\sqrt{2}\Delta^2_\mathrm{21,err}(k_i)}\right) \right],
\end{split}
\end{eqnarray}
where $\mathrm{erf}(x)$ is the error function. 
The power in various $k$ bins can be correlated due to the non-Gaussian nature of the 21 cm signal \citep[see e.g.,][]{mondal2015effect}. However, the current observation is noise dominated and it is not sensitive to the non-Gaussianity of the signal. The finite size of LOFAR stations will affect the {\it uv} tracks and therefore correlate the data in various $k$-bins. However, this effect is minor as the widths of the k-bins are large enough to minimize the correlation between the bins. Thus
the likelihood calculated above is mutually exclusive in each $k$ bin.
Therefore the total likelihood is the product of the likelihoods at various $k$ bins where we have observations. The likelihood of a parameter value $\theta$ is $\left(1-\prod_{i} \mathcal{L}(\theta|\Delta^2_\mathrm{o}(k_i))\right)$.

\section{Robustness of the constraints on the IGM parameters}
\label{appendix:igm}

\begin{figure*}
\begin{center}
\includegraphics[scale=0.62]{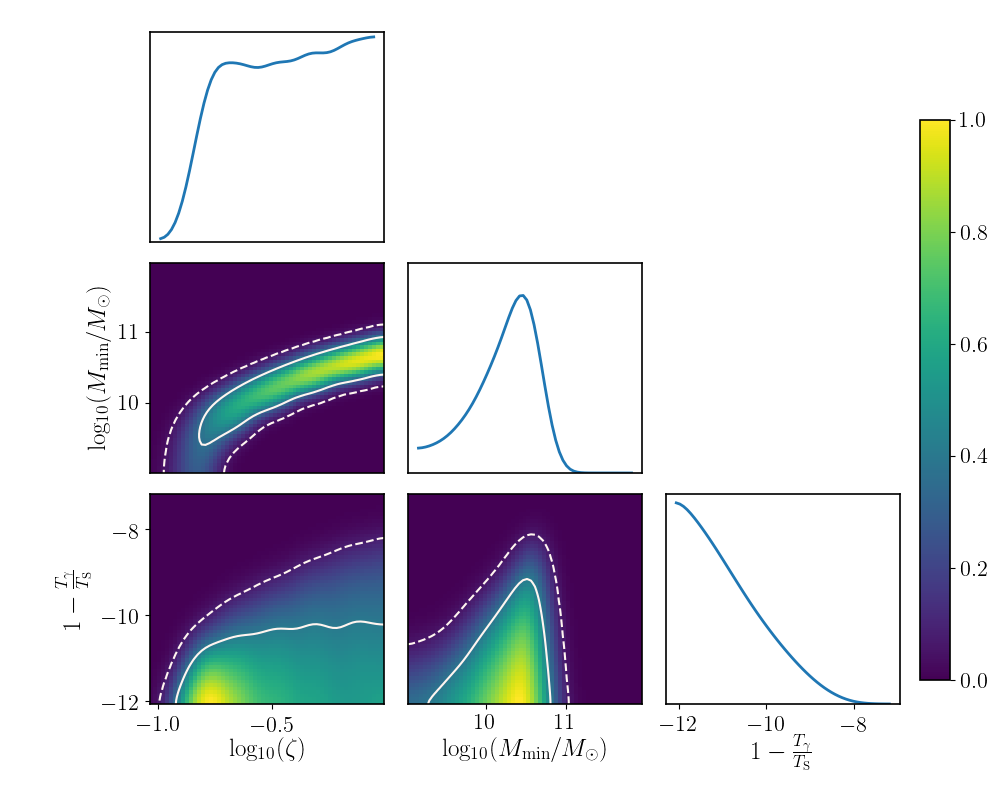}
    \caption{Similar to Fig. \ref{image_mcmcsambit} but for a different source model. Here we consider $\dot{N_i} \propto M^{\beta}_{\rm halo}$ with $\beta=1.2$.} 
   \label{image_mcmcbeta}
\end{center}
\end{figure*}

\begin{figure*}
\begin{center}
\includegraphics[scale=0.7]{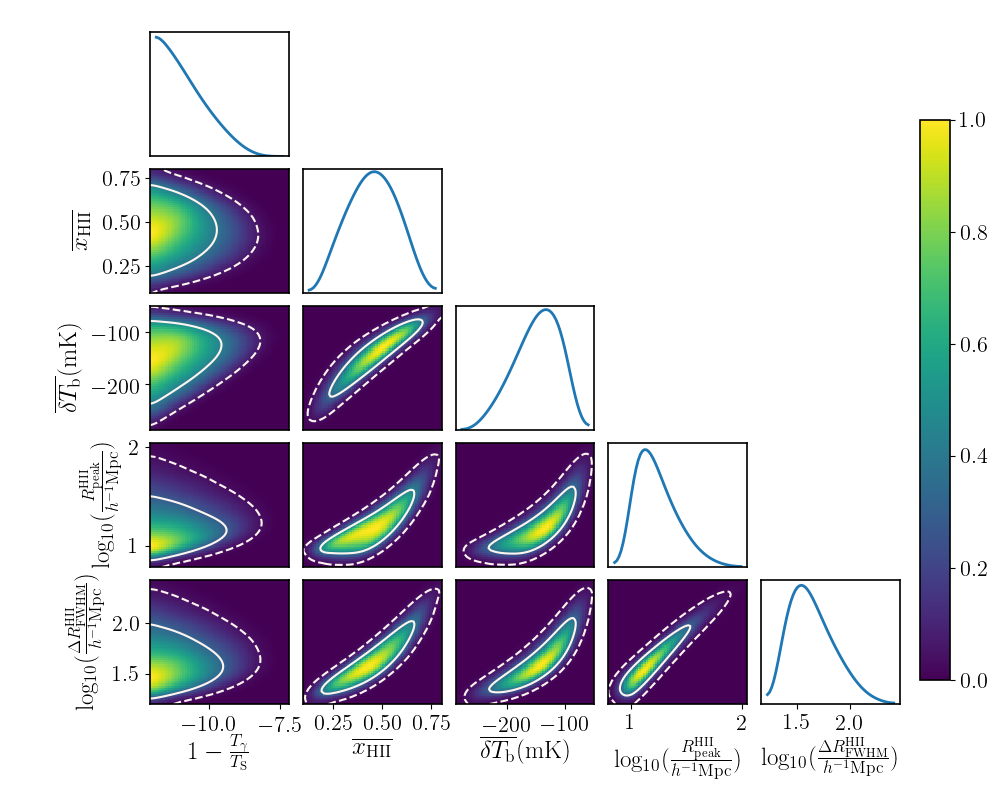}
    \caption{Similar to Fig. \ref{image_igmparam} but for a different source model. Here we consider $\dot{N_i} \propto M^{\beta}_{\rm halo}$ with $\beta=1.2$.} 
   \label{image_mcmcigmbeta}
\end{center}
\end{figure*}

We presented the constraints on the IGM parameters as our main results as the source parameters are model dependent. However, one may worry that the derived IGM parameters could somehow depend on the chosen source model. In this appendix we consider a different source model to show the robustness of the constraints on the IGM parameters. We consider the results for the uniform $\TS$ model as presented in Section \ref{sec:mcmc_units}. The original source model assumed a linear relation between stellar and halo mass: $M_\star \propto M_{\rm halo}$, see Section \ref{sec:grizzly}. Here we instead use $M_\star \propto M^{\beta}_{\rm halo}$ where we choose $\beta=1.2$, keeping the normalization constant the same as before. This source model implies that higher mass halos contribute relatively more to ionization than in the original source model.

We follow the same method as described in Section \ref{sec:emulator} to develop separate emulators for this source model using 442 simulations and explore the same parameter space as in Section \ref{sec:mcmc_units}. We also use the same number of walkers and steps in the MCMC analysis. The constraints on the source parameters and the IGM parameters from the MCMC analysis are shown in Fig. \ref{image_mcmcbeta} and \ref{image_mcmcigmbeta} respectively. 

The constraints on the source parameters $\zeta$ and $M_{\rm min}$ obviously differ from the ones shown in Section \ref{sec:mcmc_units}. As the star formation rate in the modified source model is higher compared to the original one, the part of parameter space that is ruled out shifts towards lower $\zeta$ values. However, the constraints on the IGM parameters as shown in Fig.~\ref{image_mcmcigmbeta} remain similar to what we previously found (see Fig \ref{image_igmparam}). This shows that the constraints on the IGM parameters as obtained from the upper limit observation are indeed independent of the source model as expected as the LOFAR observations do not directly measure any source properties.

\bsp	
\label{lastpage}
\end{document}